\documentclass{aa}  

\usepackage{graphicx}
\usepackage{txfonts}
\usepackage{lipsum}
\usepackage{lscape}             % to rotate a single page table, example in appendix.
\usepackage{placeins}           % useful with \FloatBarrier, to keep 
\usepackage{hyperref}
\usepackage{breakurl}
\newcommand{\urlwofont}[1]{\urlstyle{same}\url{#1}}

\usepackage{upgreek}

\def\mnras{MNRAS}
\def\apj{ApJ}
\def\apjl{ApJ}
\def\aj{AJ}
\def\araa{ARA\&A}
\def\aaps{A\&AS}
\def\aap{A\&A}
\def\pasp{PASP}
\def\apjs{ApJS}
\def\nat{Nature}
\def\procspie{SPIE Conf. Ser.}
\def\iaucirc{IAU Circ.}

\def\BaII{Ba~{\sc ii}}
\def\CaII{Ca~{\sc ii}} 
\def\HI{H~{\sc i}}
\def\HII{H~{\sc ii}}
\def\HeI{He~{\sc i}}
\def\FeII{Fe~{\sc ii}}
\def\OI{O~{\sc i}}
\def\NII{N~{\sc ii}}
\def\NaI{Na~{\sc i}}
\def\NaID{Na~{\sc i}~D}
\def\ScII{Sc~{\sc ii}}
\def\TiII{Ti~{\sc ii}}
\def\SiII{Si~{\sc ii}}

\begin{document}

\title{Explosion sites of SN 1994W-like transients}

\author{E.~Kankare\inst{1,2}\fnmsep\thanks{Corresponding author; \email{erkki.kankare@utu.fi}}, T.~Kangas\inst{3,1}, M.~Fraser\inst{4}, S.~Mattila\inst{1,5}, A.~Pastorello\inst{6}, N.~Elias-Rosa\inst{6,7}, G.~Altavilla\inst{8,9}, S.~Benetti\inst{6}, R.~Kotak\inst{1}, K.~Matilainen\inst{1}, I.~M\"antynen\inst{1}}

\institute{Department of Physics and Astronomy, University of Turku, 20014 Turku, Finland
\and Turku Collegium for Science, Medicine and Technology, University of Turku, 20014 Turku, Finland
\and Finnish Centre for Astronomy with ESO (FINCA), 20014 University of Turku, Finland
\and School of Physics, O'Brien Centre for Science North, University College Dublin, Belfield, Dublin 4, Ireland
\and School of Sciences, European University Cyprus, Diogenes street, Engomi, 1516 Nicosia, Cyprus
\and INAF -- Osservatorio Astronomico di Padova, vicolo dell'Osservatorio 5, 35122 Padova, Italy
\and Institute of Space Sciences (ICE, CSIC), Campus UAB, Carrer de Can Magrans s/n, 08193 Barcelona, Spain
\and INAF -- Osservatorio Astronomico di Roma, Via Frascati, 33, 00078 Monte Porzio Catone (RM), Italy
\and SSDC-ASI, Via del Politecnico, snc, 00133 Roma, Italy}

\date{Received November XX, 20XX}

\titlerunning{Explosion sites of SN 1994W-like transients}
\authorrunning{E. Kankare et al.}

\abstract{We study a sample of narrow-line transients that share characteristics with the Type IIn classified supernova (SN) 1994W, a prototypical member of this class of events, via investigation of their explosion sites and spectrophotometric data. The normalised cumulative rank (NCR) method was used to compare the explosion sites of 10 events to the star-formation distributions of their host galaxies, and to the sites of different evolved massive stars. The resulting sample mean value of NCR$_{\mathrm{H}\alpha} = 0.170 \pm 0.076$ is low, while the NCR$_{\mathrm{NUV}}$ distribution is flat with a mean value of $0.488 \pm 0.084$. The NCR distribution of SN 1994W-like events is consistent with relatively low-mass red supergiants (RSGs) and, despite the small sample size, inconsistent with high-mass stars such as luminous blue variables. To explain the nature of SN 1994W-like transients, interaction between an expanding ejecta and a relatively massive circumstellar medium is likely required, with the latter possibly having been produced by a H envelope ejection via a nuclear flash event, or a luminous red nova (LRN) from a stellar merger; both channels are consistent with low-mass RSGs suggested by the NCR results. In this context, we find the early $-26$ d spectrum from light curve maximum of SN 2003G to share similarities to those of F8-type supergiant stars and LRNe. Finally, based on late-time HST imaging, we set the deepest limits for the surviving precursor of SN 2011ht to $M_{\mathrm{F438W}} > -3.8$ and $M_{\mathrm{F555W}} > -4.0$ mag. This would exclude most supergiants as a non-terminal progenitor, assuming that such a star is not completely obscured by newly formed dust.}


\maketitle

\section{Introduction}

In the commonly adopted classification scheme of supernova (SN) explosions \citep{filippenko97}, Type IIn SNe are a category of stellar transients characterised by SN-like luminosities, hydrogen lines (thus Type II, contrary to H-poor Type I SNe), and narrow (thus Type IIn) line profiles with full-width at half-maximum (FWHM) of $\lesssim$10$^{3}$~km~s$^{-1}$. However, in some cases contamination through host galaxy lines or early narrow spectral features of normal Type II SNe can result in misclassification. The narrow lines of Type IIn SNe are thought to arise from the ionised unshocked circumstellar medium (CSM) and the events to be powered by shock interaction of the SN ejecta with the CSM. By definition, core-collapse SNe (CCSNe) are terminal explosions that end the life cycles of massive stars. However, the SN nature of the Type IIn classified SN~1994W and similar events have been debated \citep[e.g.][]{dessart09}.

In this work, we study a sample of spectroscopically SN~1994W-like transients; these events have overall very similar properties. Photometrically, most of these transients show a Type IIP SN-like flat $\sim$100~d light curve plateau followed by a sharp drop in luminosity, and a subsequent tail phase with decline rates that can notably vary \citep[e.g.][]{sollerman98,kankare12}. Occasionally the SN~1994W-like events are also dubbed as Type IIn-P SNe \citep[e.g.][]{mauerhan13} where the `P' indicate the presence of a plateau. During the plateau phase, the spectra are dominated by prominent H lines, accompanied by a large number of metal lines that appear roughly at +30~d, including for example \FeII, \TiII, and \CaII\ \citep[e.g.][]{dessart09}. These spectral features have distinct narrow P~Cygni profiles with broad emission wings. The absorption minima of these P~Cygni profiles have a blueshifted velocity (hereafter P~Cygni velocity) that in the case of H$\alpha$ profiles of SN~1994W-like events appear to have typically a very narrow range of $\sim$600 to 700~km~s$^{-1}$ during the initial $\sim$100~d \citep[e.g.][]{chugai04,kankare12}. If the observations have been obtained with poor spectral resolution, these velocities can appear artificially larger, or the absorption components are not detected at all. The broad line profile wings of SN~1994W-like events likely arise from multiple electron scattering \citep{dessart09}; therefore, these are not kinematically distinct high-velocity components. Furthermore, the early spectra of these transients display low H$\alpha$/H$\beta$ ratios, which can approach $L_{\mathrm{H}\alpha}/L_{\mathrm{H}\beta} \approx 1$ \citep[e.g.][]{kankare12}. Collisional de-excitation effects can reduce the H$\alpha$/H$\beta$ ratio below the typical case B recombination coefficient at high electron densities \citep{drake80}. However, at late times, in the tail phase, the spectra show large H$\alpha$/H$\beta$ ratios, similar to many other Type IIn SNe. In the light curve tail epochs, normal Type IIP SNe have entered into the nebular phase with optically thin ejecta and show broad emission lines that typically include prominent forbidden lines such as [\OI] $\lambda\lambda$6300,6364. In the post-plateau spectra, the electron scattering wings of SN~1994W-like events have disappeared, but broad nebular emission lines remain absent \citep[e.g.][]{chugai04,kankare12}. A compilation of radio observations of CCSNe by \citet{bietenholz21} covered a few SN 1994W-like events; however, those only resulted in upper limits for these transients.

Various origin channels have been discussed to explain the nature of SN~1994W-like events, including fallback SNe \citep{sollerman98,mauerhan13,elias-rosa24}, electron-capture SNe \citep{kankare12,mauerhan13,moriya14}, SN impostors via collisions of non-terminal stellar outburst shells \citep{dessart09} or specific outflow configurations \citep{humphreys12}, SNe with high CSM mass $M_{\mathrm{CSM}} \geq M_{\mathrm{ejecta}}$ \citep{dessart10,dessart16,kankare12}, or SNe with other specific ejecta and CSM properties \citep{sollerman98,chugai04,roming12,li22,elias-rosa24}.

Characteristics of SN~1994W-like events suggest that a high CSM mass compared to the ejecta mass would be required to explain these events. A strong mass-loss episode could happen via envelope ejection for which nuclear flashes or binary system mergers are potential channels. The binding energies are low, $\sim$10$^{47}$~erg, for the extended H envelopes of red supergiants (RSGs) with main-sequence masses up to $\sim$25~$M_{\sun}$ \citep{dessart10}. It has been suggested that neon or silicon flashes of $\sim$10~$M_{\sun}$ stars could lead to the ejection of the H envelope in time scales of a few years or less before a CCSN explosion \citep{woosley02,woosley15}. Based on numerical simulations of 10 to 25~$M_{\sun}$ stars, \citet{dessart10} suggested that a shell-burning-associated nuclear flash of such stars could result in an envelope ejection event that resembles a faint Type IIP SN or a SN impostor with a distinct light curve plateau, which could be a channel for a massive eruption of CSM within a few years before the SN explosion to explain events like SN~1994W.

Luminous red novae (LRNe) are transients that show double-peaked light curves that reach magnitudes of $-$13 to $-$15~mag. Typically, the first peak evolves rapidly and is blue in colour, whereas the second peak is red, slower, and in some cases plateau-like. The spectra of the events show narrow P~Cygni Balmer lines, \FeII\ features, and other metal lines. Several case studies of such events have been carried out \citep{smith16,blagorodnova17,blagorodnova21,lipunov17,macleod17,cai19,cai22,pastorello19a,pastorello19b,pastorello21a,pastorello21b,pastorello23,stritzinger20}. \citet{ivanova13} and \citet{metzger17} have discussed the contextual similarity of LRNe and fainter red novae, suggesting that LRNe arise from contact binary systems that experience a common envelope event. The pre-outburst mass loss would occur primarily in an equatorial direction powered by the conservation of angular momentum of the secondary star engulfed in the envelope, decreasing the orbital separation and resulting in a slow light curve rise \citep[e.g. $\sim$200~d in the case of the Galactic red nova V1309~Sco;][]{tylenda11}, which can be followed by a complete common envelope ejection or a merger with partial mass ejection, resulting in a double-peaked transient. \citet{metzger17} suggested a model where the first blue peak of red novae arise from thermal radiation of a hot and cooling symmetrical outburst ejecta seen unobscured in the polar direction. \citet{macleod17} and \citet{matsumoto22} have suggested that H recombination powers the red peak, though it has also been suggested that interaction with potentially equatorial low-velocity CSM can dominate or contribute to the powering of this phase \citep[e.g.][]{metzger17,matsumoto22}. The common envelope events could also be a mechanism to produce the sufficiently large mass of CSM required to explain the SN~1994W-like events \citep[e.g.][]{ercolino24}.

Section~\ref{sect:stat} describes the pixel statistics analysis carried out on the correlation of a sample of SN~1994W-like transients and the host galaxy H$\alpha$ and ultraviolet (UV) emission. Section~\ref{sect:99eb} covers the new follow-up data of SN~1999eb. Section~\ref{sect:03G} explores the archival data of SN~2003G that includes an early $-26$~d spectrum of the event. We note that the redshift time-dilation effects for our nearby sample of events would be typically quite small ($\lesssim$1~d) for the considered epochs, which are provided in the observer frame unless otherwise noted. Section~\ref{sect:HST} contains an inspection of late-time \textit{Hubble} Space Telescope (HST) archival images of the explosion site of SN~1994W-like events. Sections~\ref{sect:discussion} and \ref{sect:conclusions} provide the discussion and the conclusions, respectively.

\section{Statistical analysis of SN 1994W-like events}
\label{sect:stat}

\subsection{Sample of SN 1994W-like events}
\label{sect:sample}

\begin{figure*}
\centering
\includegraphics[width=\linewidth]{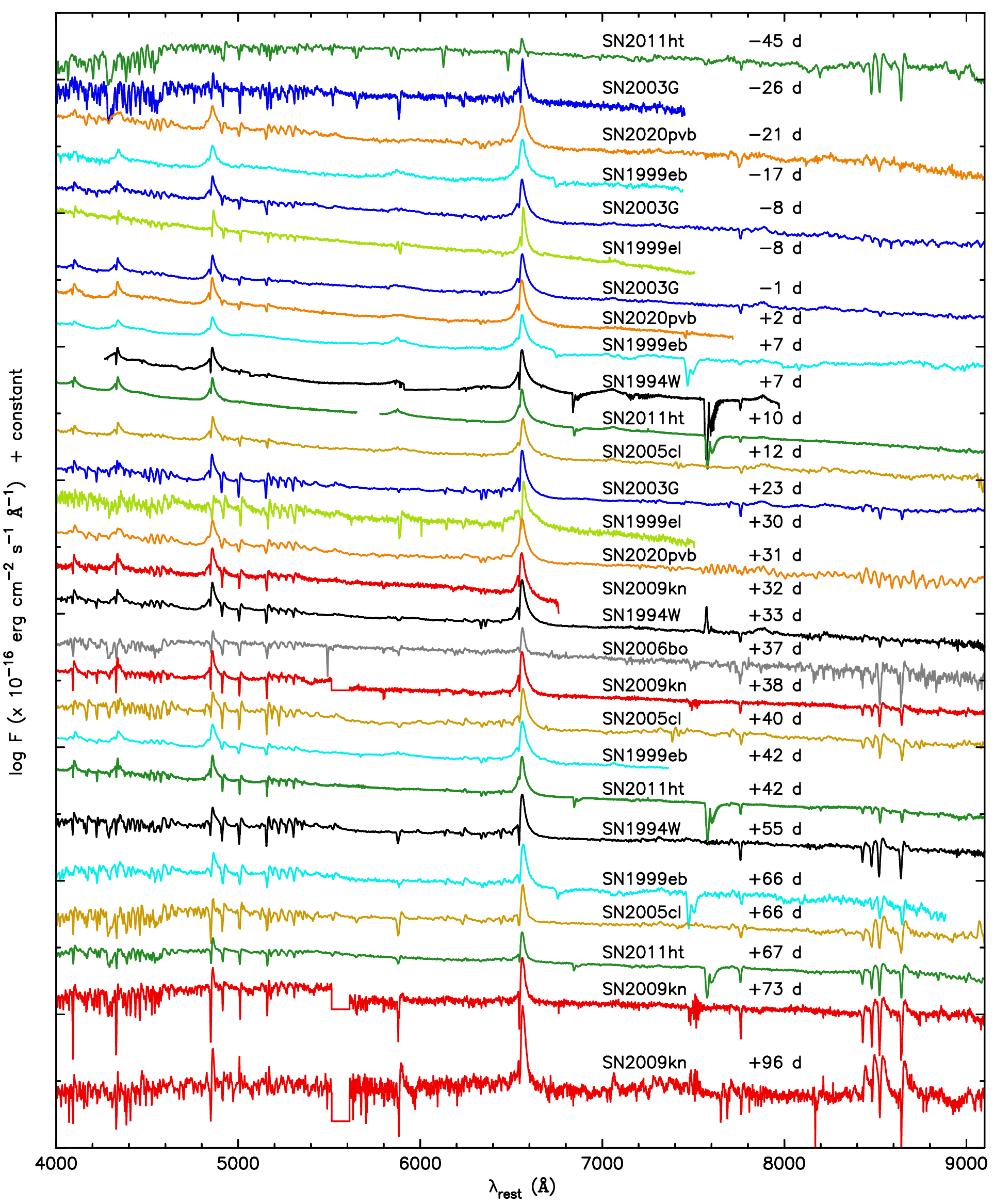}
\caption{A selection of spectra of a sample SN~1994W-like events, including SNe~1994W \citep{chugai04}, 1999eb (see Sect.~\ref{sect:99eb}), 1999el (see Sect.~\ref{sect:99el}), 2003G \citep{shivvers17}, 2005cl \citep{kiewe12}, 2006bo \citep{taddia13}, 2009kn \citep{kankare12}, 2011ht \citep{humphreys12,pastorello19b}, and 2020pvb \citep{elias-rosa24}. The spectra were dereddened by the total line-of-sight extinction estimates and the wavelengths corrected to the host galaxy rest frame.}
\label{fig:sample}
\end{figure*}

\begin{figure}
\centering
\includegraphics[width=\linewidth]{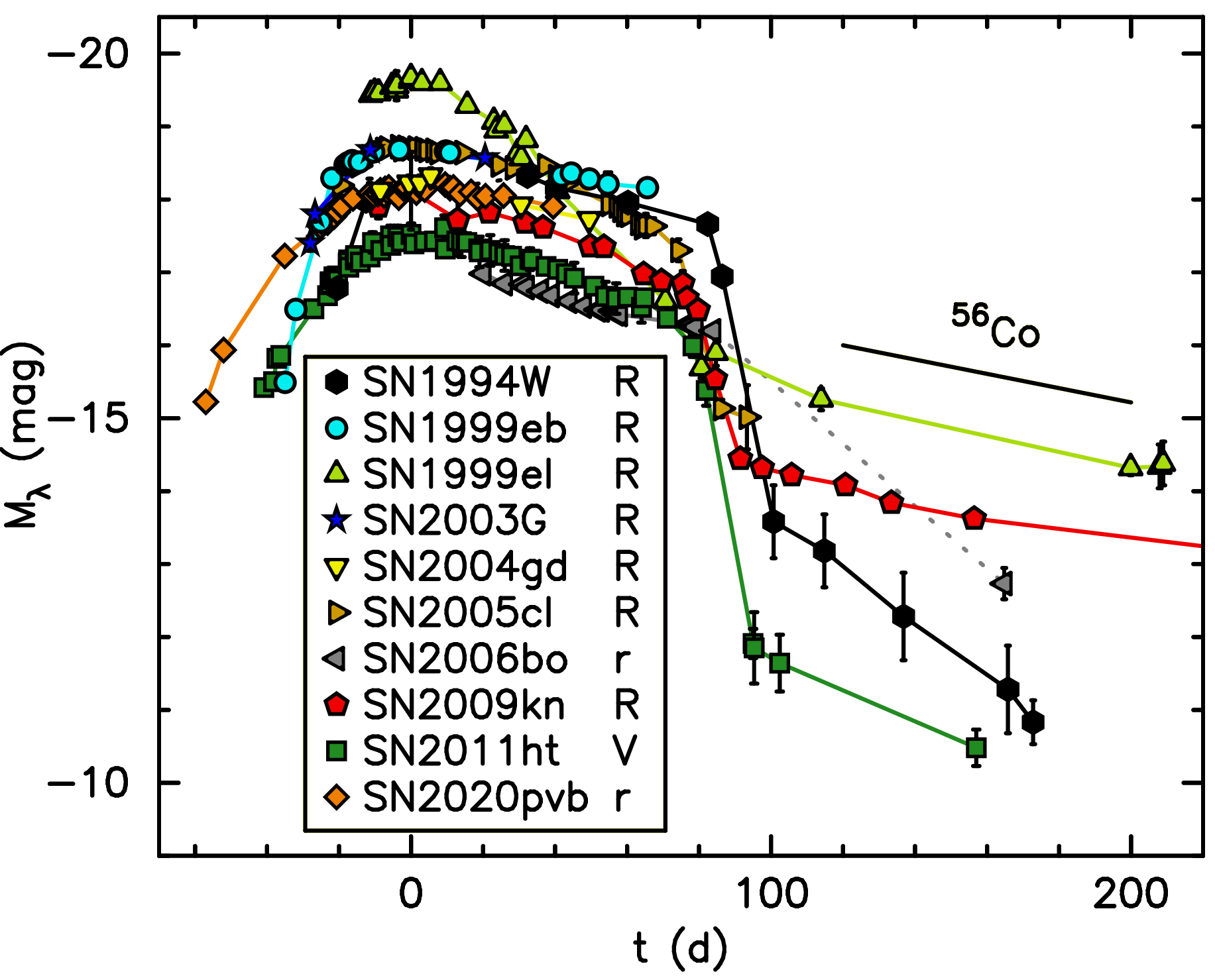}
\caption{Absolute light curve selection of spectroscopically SN~1994W-like events. The epoch $t = 0$~d corresponds to the estimated light curve peak. The sample includes SNe~1994W \citep[][and references therein]{sollerman98}, 1999eb (see Sect.~\ref{sect:99eb}), 1999el \citep{dicarlo02}, 2003G (see Sect.~\ref{sect:03G}), 2004gd (see Sect.~\ref{sect:03G}), 2005cl \citep{kiewe12}, 2006bo \citep{taddia13}, 2009kn \citep{kankare12}, 2011ht \citep{roming12,humphreys12,mauerhan13}, and 2020pvb \citep{elias-rosa24}. The \textit{R}-band light curve of SN~1994W was partly generated based on the \textit{B} and \textit{V}-band observations of the event, and the colour evolution of SN~2009kn. The luminosity slope is shown for the radioactive decay of $^{56}$Co assuming complete $\gamma$-ray trapping.}
\label{fig:lc}
\end{figure}

This section lists the SN~1994W-like events adopted in our analysis, which likely share a similar physical origin. A spectroscopic time-series is shown in Fig.~\ref{fig:sample} with a selection of spectra of some of these transients to stress their homogenous evolution. Similarly, a light curve comparison is shown in Fig.~\ref{fig:lc}. The basic parameters of the sample transients are presented in Table~\ref{table:sample}. Preceding our observations for the statistical study, the sample was selected based on classified Type IIn SNe that were stated in telegrams or journal publications to spectroscopically resemble SN~1994W or other events previously identified as SN~1994W-like. Further details on these reports are provided in the following subsections for each target. Unfortunately, classification descriptions do not necessarily explicitly state similarities to SN~1994W-like events or the observations are not obtained with sufficiently high resolution and signal-to-noise ratio to identify SN~1994W-like characteristics. Therefore, our sample is likely missing potential events; however, those should not be particularly biased towards any specific environments.

\begin{table*}
\caption{Adopted parameters of the analysed sample of SN~1994W-like events.}
\centering
\begin{tabular}{lccccccccc}
\hline
\hline
SN & $\alpha$\tablefootmark{a} & $\delta$\tablefootmark{a} & Host & $z$\tablefootmark{b} & $D_{L}$\tablefootmark{c} & $\mu$ & $A_{V,\mathrm{Gal}}$\tablefootmark{d} & $A_{V,\mathrm{host}}$\tablefootmark{e} & $t_{\mathrm{peak,JD}}$ \\
 & & & & & (Mpc) & (mag) & (mag) & (mag) & \\
\hline
1994W & 12:02:11.01 & +62:08:31.8 & NGC 4041 & 0.004116 & 24.4 & 31.94 & 0.049 & - & 2449571.5 \\
1994ak & 09:14:01.47 & +40:06:21.5 & NGC 2782 & 0.008483 & 41.5 & 33.09 & 0.044 & - & - \\
1999eb & 01:43:45.45 & +04:13:25.9 & NGC 664 & 0.018096 & 76.8 & 34.43& 0.074 & 0.0 & 2451475.9 \\
1999el & 20:37:17.72 & +66:06:11.5 & NGC 6951 & 0.004750 & 25.5 & 32.03 & 1.020 & 1.5 & 2451490.6 \\
2003G & 02:08:28.13 & +06:23:51.9 & IC 208 & 0.011755 & 49.5 & 33.47 & 0.149 & 0.0 & 2452675.7 \\
2004F & 03:17:53.80 & $-$07:17:43.0 & NGC 1285 & 0.017475 & 74.1 & 34.35 & 0.150 & - & - \\
2004gd & 07:09:11.71 & +20:36:10.6 & NGC 2341 & 0.017215 & 76.4 & 34.42 & 0.198 & 1.0 & 2453324.5 \\
2005cl & 21:02:02.35 & $-$06:17:35.7 & MCG -01-53-020 & 0.025878 & 116.0 & 35.32 & 0.194 & - & 2453556.5 \\
2006bo & 20:30:41.90 & +09:11:40.8 & UGC 11578 & 0.015347 & 70.6 & 34.24 & 0.283 & - & 2453814.9 \\
2009kn & 08:09:43.04 & $-$17:44:51.3 & ESO 561- G 020 & 0.015798 & 71.8 & 34.28 & 0.306 & 0.0 & 2455140.0 \\
2011ht & 10:08:10.59 & +51:50:57.0 & UGC 5460 & 0.003646 & 21.4 & 31.65 & 0.029 & 0.19 & 2455879.5 \\
2020pvb & 20:53:53.03 & $-$25:28:26.1 & NGC 6993 & 0.020277 & 90.3 & 34.78 & 0.187 & 0.15 & 2459159.8 \\ 
\hline
\end{tabular}
\tablefoot{
\tablefoottext{a}{SN coordinates from the Transient Name Server (TNS) except SN~1994W (see Sect.~\ref{sect:94W}).}
\tablefoottext{b}{Redshifts via the NASA/IPAC Extragalactic Database (see references therein).}
\tablefoottext{c}{Virgo + Great Attractor corrected luminosity distance \citep{mould00} assuming $H_0 = 69.6$~km~s$^{-1}$~Mpc$^{-1}$ \citep{bennett14}.}
\tablefoottext{d}{Galactic line-of-sight extinction values from \citet{schlafly11}.}
\tablefoottext{e}{Host extinction estimates from the literature for SNe~2009kn \citep{kankare12}, 2011ht \citep{roming12}, 2020pvb \citep{elias-rosa24}, or from this work for SNe~1999eb, 1999el, 2003G, and 2004gd. Otherwise, negligible host galaxy extinction was assumed.}
}
\label{table:sample}
\end{table*}

\subsubsection{SN 1994W}

SN~1994W was discovered by \citet{cortini94} in NGC~4041 on Julian date (JD) 2449563.35. The event was classified as a Type II SN with narrow Balmer lines based on observations on JD~2449565.4 by \citet{bragaglia94} and on JD~2449568.5 by \citet{filippenko94}, with the latter reporting P~Cygni velocities of 700~km~s$^{-1}$. In their spectrophotometric study, \citet{sollerman98} estimated a $^{56}$Ni mass of $\lesssim$0.0026~$M_{\odot}$ and suggested that the event was either a low-energy explosion of a massive star with a low zero age main sequence (ZAMS) mass ($M_\mathrm{ZAMS} \approx 8$ to $10$~$M_{\sun}$), or a fallback SN from a high-mass ($M_\mathrm{ZAMS} \gtrsim 25$~$M_{\sun}$) progenitor, with the latter scenario favoured. \citet{chugai04} carried out modelling of the light curves and H$\alpha$ line profile evolution of SN~1994W, and suggested that the event was a Type IIn SN with an ejecta mass of around 8~$M_{\sun}$ embedded in a 0.4~$M_{\sun}$ envelope of CSM. However, \citet{dessart16} pointed out based on further hydrodynamical and radiative transfer simulations that such a scenario would struggle not to produce broad lines at late times, which were absent in SN~1994W, and would instead result in a SN~1998S-like event \citep{fassia01}. \citet{dessart09} carried out radiative transfer modelling of SN~1994W and found that the plateau phase line profiles, with broad wings included, can be produced by electron scattering via a single region of radiation in a slowly expanding material; based on the absence of broad emission line components and low $^{56}$Ni mass estimates they also speculated with the possibility that the event could have been powered by a collision of ejected shells from non-terminal stellar explosions. Based on their simulations, \citet{dessart16} favoured a model of a fast moving but less massive SN ejecta interacting with a more massive CSM shell ($M_{\mathrm{CSM}} > M_{\mathrm{ejecta}}$) to explain the observational characteristics of SN~1994W. 

\subsubsection{SN 1994ak}

\citet{nakano94} reported the discovery of SN~1994ak in NGC~2782 with the earliest detection on JD~2449711.36. A Type IIn classification spectrum of SN~1994ak was obtained by \citet{garnavich95} on JD~2449714.95 and reported that it showed strong Balmer lines with a narrow emission component on a broad base, and P~Cygni velocities of 600~km~s$^{-1}$. No detailed analysis of SN~1994ak has been published; however, \citet[][see their figure 14]{filippenko97} presented in the review of CCSN subtypes an optical spectrum of SN~1994ak obtained on JD~2449744, which shows narrow P~Cygni Balmer and \FeII\ lines similar to SN~1994W-like events. 

\subsubsection{SN 1999eb}

\citet{modjaz99} reported the discovery of SN~1999eb in NGC~664 using the 0.8~m Katzman Automatic Imaging Telescope (KAIT) with the earliest unfiltered 19.0~mag detection on JD~2451440.5 and three additional epochs of early photometry that showed a light curve rise. We adopted these observations as approximate \textit{R}-band magnitudes. \citet{garnavich99} obtained a spectrum of the event on JD~2451454.8, and reported Balmer lines with P~Cygni velocities of 900~km~s$^{-1}$, P~Cygni \FeII\ lines, and an overall similarity to that of SN~1995G\footnote{SN~1995G showed spectroscopic similarity to SN~1994W-like events with P~Cygni profile in Balmer and \FeII\ lines that have blueshifted absorption minimum velocities of $\sim$700~km~s$^{-1}$; however, the event has also a very slow light curve evolution of $\sim$5~mag in 2.5~yr \citep{pastorello02}. It is unclear if SN~1995G shares the origin channel with other SN~1994W-like transients discussed here, and conservatively it was not included in our sample. Similarly, while the early spectra of SN~2011A \citep{dejaeger15} resemble SN~1994W-like events, the later spectral evolution differ and the event was not included in the sample.}. \citet{li02} presented HST late-time F555W and F814W band observations of SN~1999eb, which we adopted as approximate \textit{V} and \textit{I}-band magnitudes, respectively (Sect.~\ref{sect:99eb}).

\subsubsection{SN 1999el}
\label{sect:99el}

SN~1999el was discovered in NGC~6951 on JD~2451471.95. \citet{cao99} reported the discovery and classification of the event as a Type IIn SN. \citet{dicarlo02} carried out a spectrophotometric follow-up study of the transient and concluded that the event was a Type IIn SN with a rapidly declining light curve, which showed narrow Balmer lines with broad wings and P~Cygni velocities of 600 to 900~km~s$^{-1}$; the spectroscopic similarity to that of SN~1999eb was pointed out in the study. They suggested that SN~1999el had a progenitor with a history of asymmetric mass-loss episodes and a near-infrared (near-IR) excess that arose likely from pre-existing circumstellar dust. Similar to SN~1999eb, \citet{li02} reported HST late-time F555W and F814W band observations of SN~1999el. Spectroscopically, SN~1999el is very similar to SN~1994W-like transients, and has a linear Type IIL SN-like light curve evolution; therefore, a distinctive Type IIP SN-like plateau and a subsequent light curve drop are not necessarily characteristics that define this type of event.

One spectrum of SN~1999el was reported by \citet{shivvers17} in their compilation of SN data, and the Weizmann Interactive Supernova Data Repository \citep[WISeREP;][]{yaron12} hosts six additional epochs of spectra with no available information on the used instruments. The host galaxy extinction of SN~1999el is uncertain; therefore, we carried out a comparison between the spectrum in WISeREP of SN~1999el observed +30~d after the light curve peak and dereddened with a range of different host galaxy extinctions, and the spectrum of SN~2009kn which was deemed to have been obtained roughly +32~d after the light curve peak. This suggests a host galaxy extinction of $A_{V,\mathrm{host}} \approx 1.5$~mag to result in similar colours (Fig.~\ref{fig:sample}). Combined with the Galactic extinction of $A_{V,\mathrm{Gal}} = 1.020$~mag this is consistent with the high-end estimate of \citet{dicarlo02} with an accompanying $^{56}$Ni mass estimate of $\leq$0.07~$M_{\odot}$.

\subsubsection{SN 2003G}
\label{sect:SN2003G}

\citet{graham03} reported the discovery of SN~2003G in IC~208 on JD~2452647.7 with an unfiltered magnitude of 16.2~mag, and an additional imaging on JD~2452649.0 with 15.8~mag, which we adopt directly as \textit{R}-band magnitudes. The classification spectrum was obtained by \citet{hamuy03} on JD~2452648.55, which they note to show strong similarity to SN~1994W with line profiles that have narrow emission and absorption components and broad bases. No detailed study of the event has been reported; however, four spectra are available in WISeREP, which we adopted to our analysis (Sect.~\ref{sect:03G}).

\subsubsection{SN 2004F}

\citet{swift04} reported the discovery of SN~2004F in NGC~1285 on JD~2453020.7 with an unfiltered magnitude of 17.8~mag. The reported unfiltered VSNET magnitude\footnote{\urlwofont{http://www.astrosurf.com/snweb2/2004/04F_/04F_Meas.htm}} of 17.1~mag on JD~2453029.34 suggests that the light curve was rising and corresponds to an absolute magnitude of roughly $-17.4$~mag. \citet{filippenko04a} reported classification spectra of SN~2004F obtained around JD~2453049, and noted that SN~2004F resembles SNe~1994W and 1994ak with narrow emission lines and P~Cygni velocities of 800~km~s$^{-1}$. They noted that \CaII, \OI, \FeII, and \NaI\ features were present in the spectrum. 

\subsubsection{SN 2004gd}

\citet{moore04} reported the discovery of SN~2004gd in NGC~2341 on JD~2453315.95 with an unfiltered magnitude of 18.3~mag, and three additional follow-up epochs that cover a slow light curve rise. We adopted these observations as approximate \textit{R}-band magnitudes. \citet{filippenko04b} obtained a spectrum of SN~2004gd on JD~2453352, and reported that it resembled SN~1995G with narrow emission lines on a broad base and P~Cygni velocities of 700~km~s$^{-1}$. 

\subsubsection{SN 2005cl}

SN~2005cl was discovered in MCG~-01-53-020 by \citet{pugh05} with the earliest detection on JD~2453523.97. \citet{modjaz05} reported a spectroscopic classification on JD~2453534.93 as a Type IIn SNe with a blue continuum and Balmer emission lines with narrow and broad components. Spectrophotometric follow-up data of SN~2005cl was presented in a compilation of four Type IIn SNe by \citet{kiewe12}, reporting narrow P~Cygni profiles of Balmer lines and an overall spectrophotometric similarity to that of SN~1994W. Their optical photometry covered a plateau length of roughly 100~d, which turned to a more rapid decline. \citet{li22} used a radiation hydrodynamics simulation grid to model the light curve evolution of SN~2005cl, and suggested that SN~1994W-like events could be explained as $M_\mathrm{ZAMS} \approx 10$ to $15$~$M_{\sun}$ RSG progenitors that have experienced a $\sim$10$^{47}$~erg outburst several months before the final SN explosion (best fit 258~d before the onset of SN~2005cl suggested by their models). 

\subsubsection{SN 2006bo}

\citet{boles06} reported the discovery of SN~2006bo in UGC~11578 with the earliest detection on JD~2453830.65. \citet{blondin06} classified the event as a Type IIn SN based on a spectrum obtained on JD~2453851.01. The compilation of five Type IIn SNe by \citet{taddia13} included a follow-up study of SN~2006bo, which they identified as a SN~1994W-like event with Balmer and \FeII\ lines that show narrow emission components and P~Cygni velocities of 600~km~s$^{-1}$. The photometry of SN~2006bo does not cover the rise and the explosion date is uncertain as the pre-discovery non-detection on JD~2453703 does not provide a strong constraint. The photometric observations of SN~2006bo by \citet{taddia13} started on JD~2453834.92. Based on the light curve evolution and the continuum shape of the SN~2006bo spectrum obtained on JD~2453841.5, we estimated that their photometric follow-up observations started roughly 20~d after the light curve maximum. 

\subsubsection{SN 2009kn} 

\citet{gagliano09} discovered SN~2009kn in ESO~561-~G~020 on JD~2455130.96. \citet{steele09} reported the classification of SN~2009kn as a Type IIn SN based on a spectrum obtained on JD~2455145 and noted strong Balmer emission lines and narrow absorption components. \citet{kankare12} carried out a detailed follow-up study of the event, showed that it had twin-like characteristics compared to those of SN~1994W, and discussed the origin scenario of a SN with a massive CSM (including a variation of an electron-capture SN with an ambient medium of an ejected stellar envelope); the late-time photometry was found to be roughly consistent with a radioactive decay tail with a $^{56}$Ni mass estimate of 0.023~$M_{\odot}$. Based on light curve modelling, \citet{moriya14} suggested that an electron-capture SN from a super-AGB star could produce a $-17$~mag transient with a H recombination powered plateau somewhat similar to SN~2009kn.

\subsubsection{SN 2011ht}

\citet{boles11} discovered SN~2011ht in UGC~5460 on JD~2455833.68. \citet{pastorello11} reported a spectroscopic classification of the event on JD~2455834.72 and considered the event to be a SN impostor, noted narrow P~Cygni spectral line profiles, and listed detections of \HI, \CaII, \FeII, \NaID, \ScII, and \BaII. Early \textit{Swift} observations of the event were reported by \citet{roming11}. \citet{prieto11} classified the event as a Type IIn SN based on a spectrum obtained on JD~2455877.0, and reported Balmer lines with narrow and broad emission components, and P~Cygni velocities of 700~km~s$^{-1}$. \citet{roming12} noted that SN~2011ht showed dense stellar wind-like spectral characteristics that resembled some SN impostors, but had the energy of a terminal SN explosion. They discussed the possibility of a dense shell of CSM ejected $\sim$1~yr before the SN and shocked by the subsequent explosive transient. \citet{humphreys12} noted that observationally SN~2011ht was remarkably similar to that of SN~1994W, the inferred radiated energy of SN~2011ht was relatively low, and a likely origin for the event would be a continuous outflow with an onset $\sim$0.5~yr before the discovery followed by outflow acceleration or a high-velocity mass-loss episode. However, \citet{mauerhan13} argued that a wind-like outflow would require an unphysically high Eddington factor of $\sim$1000 to explain the event. They also found the tail phase of SN~2011ht to be consistent with radioactive decay of $^{56}$Co when the near-IR excess associated with dust formation was included in the pseudo-bolometric light curve, resulting in a low estimated $^{56}$Ni mass range of 0.006 to 0.01~$M_{\sun}$, and suggested the event to be a terminal SN explosion, possibly either from an electron-capture SN or from a fallback SN, interacting with a dense CSM. 

\citet{fraser13} and  \citet{ofek14} reported precursor outburst detections of SN~2011ht, which resulted in a combined coverage within epochs $-367$ to $-185$~d before the maximum of SN~2011ht adopted here. \citet{pastorello19b} referred to this pre-outburst transient as UGC~5460-2010OT1 and discussed the similarity of the early spectrum of SN~2011ht, at $-45$~d before our adopted light curve peak, with that of a LRN NGC~4490-2011OT1 during the red peak. Furthermore, the light curve of NGC~5460-2010OT1 shares contextual similarity to those of double-peaked LRNe. Based on these characteristics, it was suggested that the precursor of SN~2011ht could have been a merger that had powered a LRN event.

\subsubsection{SN 2020pvb}

The discovery of SN~2020pvb in NGC~6993 was reported by \citet{chambers20}. As discussed by \citet{elias-rosa24}, the early pre-outburst detection of $m_{w} = 21.04$~mag on JD~2459048.92 was followed by non-detections and a subsequent brightening to the main SN-like event with the first detection on JD~2459099.87. The transient was classified as a Type IIn SN on JD~2459138.3 by \citet{perley20}. \citet{elias-rosa24} reported the detailed study of SN~2020pvb, showed that it was a SN~1994W-like event and concluded that the transient could be a fallback SN from a massive progenitor, or a SN from a moderate or low-mass progenitor with a dense CSM.

\subsection{NCR analysis of SN 1994W-like events}
\label{sect:ncr}

\citet{fruchter06} presented the statistical pixel analysis method in a correlation study of a sample of gamma-ray bursts and their host galaxy rest-frame UV/blue light. Subsequently, the normalised cumulative rank (NCR) method has been used to study CCSN subtypes via their correlations with host galaxy emission, in particular in H$\alpha$ \citep[e.g.][]{james06,anderson12,kangas13}. Furthermore, \citet{kangas17} compared the NCR results from CCSNe to those of massive stars in nearby galaxies. We carried out an NCR analysis of a sample of 10 spectroscopically SN~1994W-like events (Sect.~\ref{sect:sample}) to constrain their progenitor mass range. SNe~2005cl and 2020pvb were not included in the NCR sample due to their large distance of 116 and 90~Mpc, respectively, to avoid systematic uncertainties \citep{kangas17}.

\subsubsection{NCR method}

In the NCR method, the pixels of a galaxy image are ranked from faintest to brightest in an ascending order. The NCR value of each pixel is defined as a fraction of the host emission that originates in pixels with lower brightness. If the bias and sky background level, and in the case of line emission (e.g. H$\alpha$) the galaxy continuum, have been accurately subtracted, the average pixel value $p$ outside the galaxy is 0. Therefore, if the sum of the ascending pixel values is $\leq$0, the NCR value is set to 0; this is the background. If the sum is $>$0, these are the regions of host galaxy emission, and the NCR value is defined as $\mathrm{NCR}(n) = \sum^{n}_{i=1} p_{i} / p_{\mathrm{tot}}$, where $p_{\mathrm{tot}}$ is the pixel value sum over the whole image and $n$ is the $n^{\mathrm{th}}$ pixel in the ascending order. The exact image size does not affect the results. The NCR values of the pixels associated with sample target coordinates are a tracer of any correlation between the targets and the underlying host emission, and can be used for statistical comparison. 

Over time, H$\alpha$ emission of a star forming region dissipates, and the stars can stray further away from their birth regions. Therefore, NCR$_{\mathrm{H}\alpha}$ associations of the various CCSN subtypes have been interpreted as a tracer of the progenitor star age sequence. Type Ic SNe are thought to originate from the most massive and short-lived progenitors, which closely trace very recent star formation, and therefore the H$\alpha$ emission (i.e. diagonal NCR$_{\mathrm{H}\alpha}$ distribution in Fig.~\ref{fig:ncr}). Other SN types present an excess in low NCR$_{\mathrm{H}\alpha}$ values; Type IIP and IIn SNe show the strongest deviation, which indicates relatively low-mass, long-lived progenitors that no longer are associated with the H$\alpha$ emission of the parent stellar population. Qualitatively, the observed trend of NCR$_{\mathrm{H}\alpha}$ statistics corresponds to the canonical understanding of the increasing progenitor mass sequence of basic CCSNe subtypes from Type IIP to Ib to Ic \citep[e.g.][]{anderson08,anderson12,kangas13}. However, as pointed out as a caveat by \citet{crowther13}, rather than probing individual \HII\ regions, typical survey pixel scales and galaxy sample distances of NCR studies cover physical scales of a few hundred pc. Such scales are dominated by larger star formation complexes, which can have long, and therefore somewhat non-constraining, lifetimes of tens of Myr. Furthermore, it is also plausible that due to projection effects massive stars with relatively low masses can be seen associated with high NCR values, or very massive stars can be hosted in isolated and faint \HII\ regions with low NCR values. However, statistically the less massive stars with longer lifetimes are nonetheless also less likely to have high NCR values. This assumption was verified by \citet{kangas17}, who found a clear correlation between NCR values of evolved massive stars and basic CCSN subtypes, consistent with the canonical picture of stellar evolution.

An intriguing NCR mapping result is that the Type IIn SN explosion sites do not follow their host galaxy H$\alpha$ emission \citep{anderson08, anderson12, habergham14, ransome22}. As shown in Fig.~\ref{fig:ncr}, in the most extensive Type IIn NCR sample of 75 events by \citet{ransome22} 50~\% of these SNe fall in regions of 0 or very faint H$\alpha$ flux (lowest 10~\% bin). However, NCR values obtained by \citet{kangas17} for example for luminous blue variables (LBVs) in the Large Magellanic Cloud (LMC) and M~33 are higher than those of Type IIn SNe; the low NCR values imply that at least part of Type IIn SNe arise from low-mass progenitors, and the low-mass RSGs were found to have the most similar distribution to those of Type IIn SNe. However, Type IIn SNe are a heterogeneous class and a variety of progenitors is plausible \citep[e.g.][]{fraser20}.

\subsubsection{NCR data}

We obtained narrow-band H$\alpha$ imaging of the host galaxy sample of SN~1994W-like events using the 2.56~m Nordic Optical Telescope \citep[NOT;][]{djupvik10} with the Alhambra Faint Object Spectrograph and Camera (ALFOSC) equipped with a 2048 $\times$ 2048 pixel detector and a plate scale of 0$\farcs$19~pixel$^{-1}$ (the CCD8 detector at the time of the observations). Our sample was limited to events at a distance of $<$80~Mpc. Our imaging was obtained using both a narrow `ON' filter that covers the H$\alpha$ line and a narrow `OFF' filter close in wavelength to cover a line-free section of the continuum to subtract that from the H$\alpha$ observations. The used filters were selected based on the redshift of the host galaxy and are reported in Table~\ref{table:ncr} with the identification convention used by the NOT. The used selection of narrow-band filters included \#21 ($\lambda_{\mathrm{eff}} = 6564$~\AA, $\mathrm{FWHM} = 33$~\AA), \#49 ($\lambda_{\mathrm{eff}} = 6610$~\AA, $\mathrm{FWHM} = 50$~\AA), \#50 ($\lambda_{\mathrm{eff}} = 6653$~\AA, $\mathrm{FWHM} = 55$~\AA), and \#52 ($\lambda_{\mathrm{eff}} = 6725$~\AA, $\mathrm{FWHM} = 60$~\AA). The centre of the H$\alpha$ line is always located within the FWHM of the ON filter. The basic image reduction, including bias subtraction and flat field correction, was carried out using Image Reduction and Analysis Facility \citep[{\sc iraf};][]{tody86} tasks, which were also used to align the ON and OFF image pairs. The image subtraction process between the ON and OFF images of the sample galaxies was carried out with a slightly modified version of the {\sc isis} 2.2 package \citep{alard98,alard00} to allow manual selection of bright field stars and their surrounding background as reference data for the algorithm to convolve the point spread function (PSF) of the higher quality image to match that of the lower quality image, and adjust the image levels. The world coordinate system was derived for the analysed images using {\sc iraf} tasks and field stars listed in the USNO-A2.0 catalogue \citep{monet98}, typically of the order of $\sim$10 sources in the ALFOSC 6$\farcs$5~$\times$~6$\farcs$5 field of view (FOV). The NCR method does not require absolute calibration of the image, as the NCR value is defined as a fraction of emission in the galaxy from pixels fainter (or equal) to that of the target coordinates. The observed H$\alpha$ emission does include some flux from the host galaxy [\NII] $\lambda\lambda$6548,6584 doublet; however, this effect is minimal for the NRC analysis \citep{kangas17,gonzalez-diaz24} and applies similarly to both transient and stellar comparison samples.

\subsubsection{NCR results}

\begin{figure*}
\centering
\includegraphics[width=0.32\linewidth]{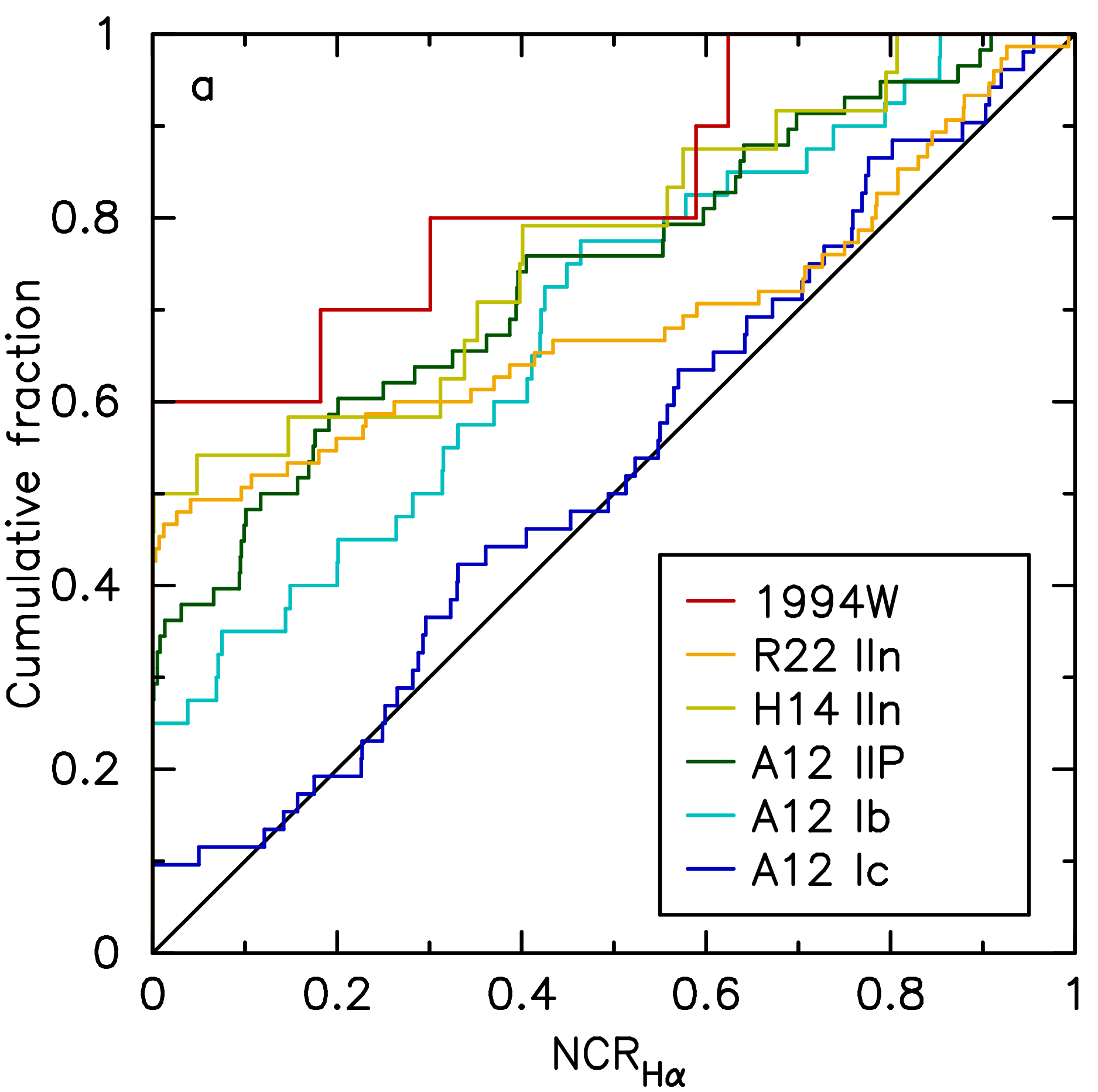}
\includegraphics[width=0.32\linewidth]{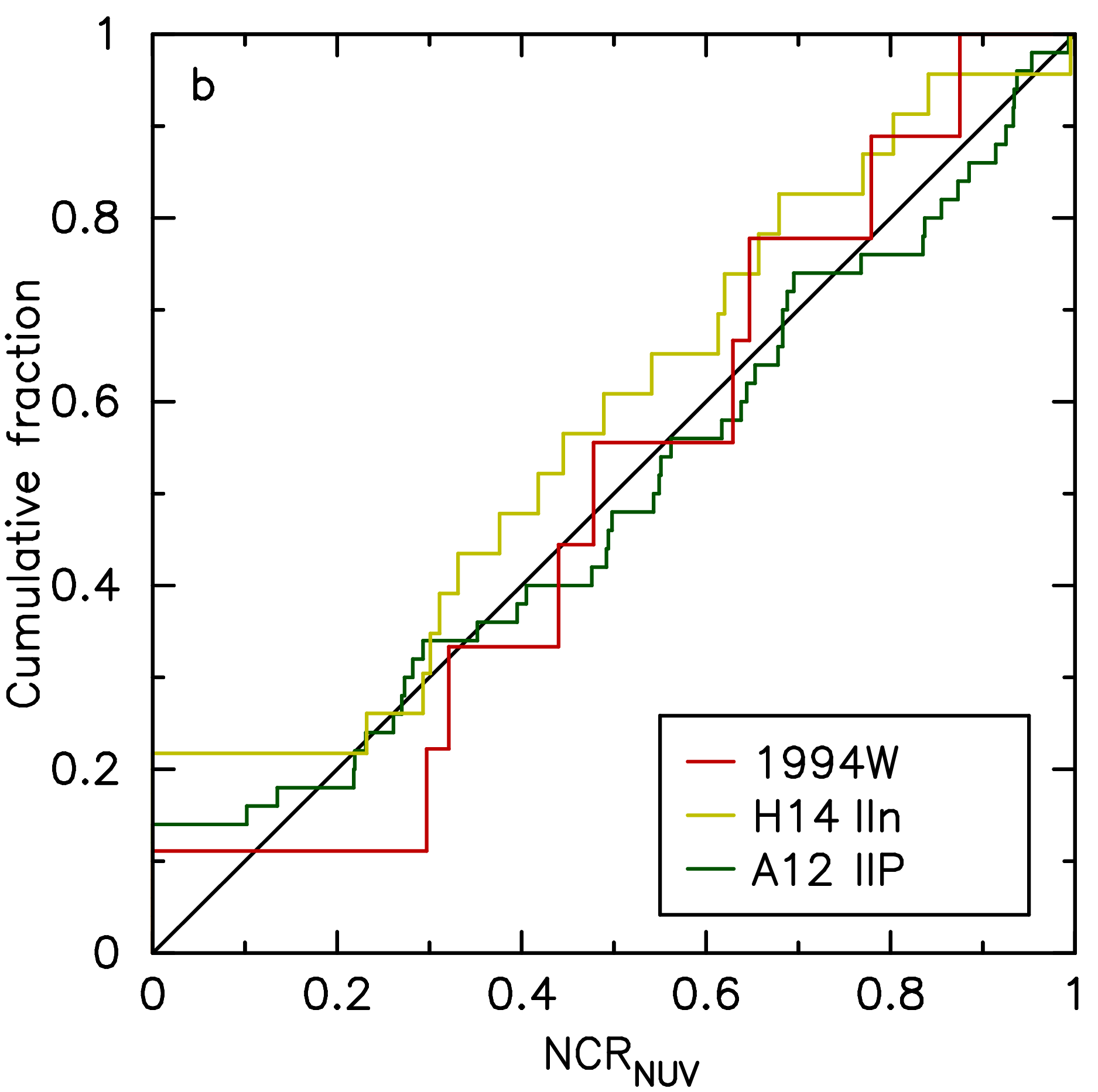}
\includegraphics[width=0.32\linewidth]{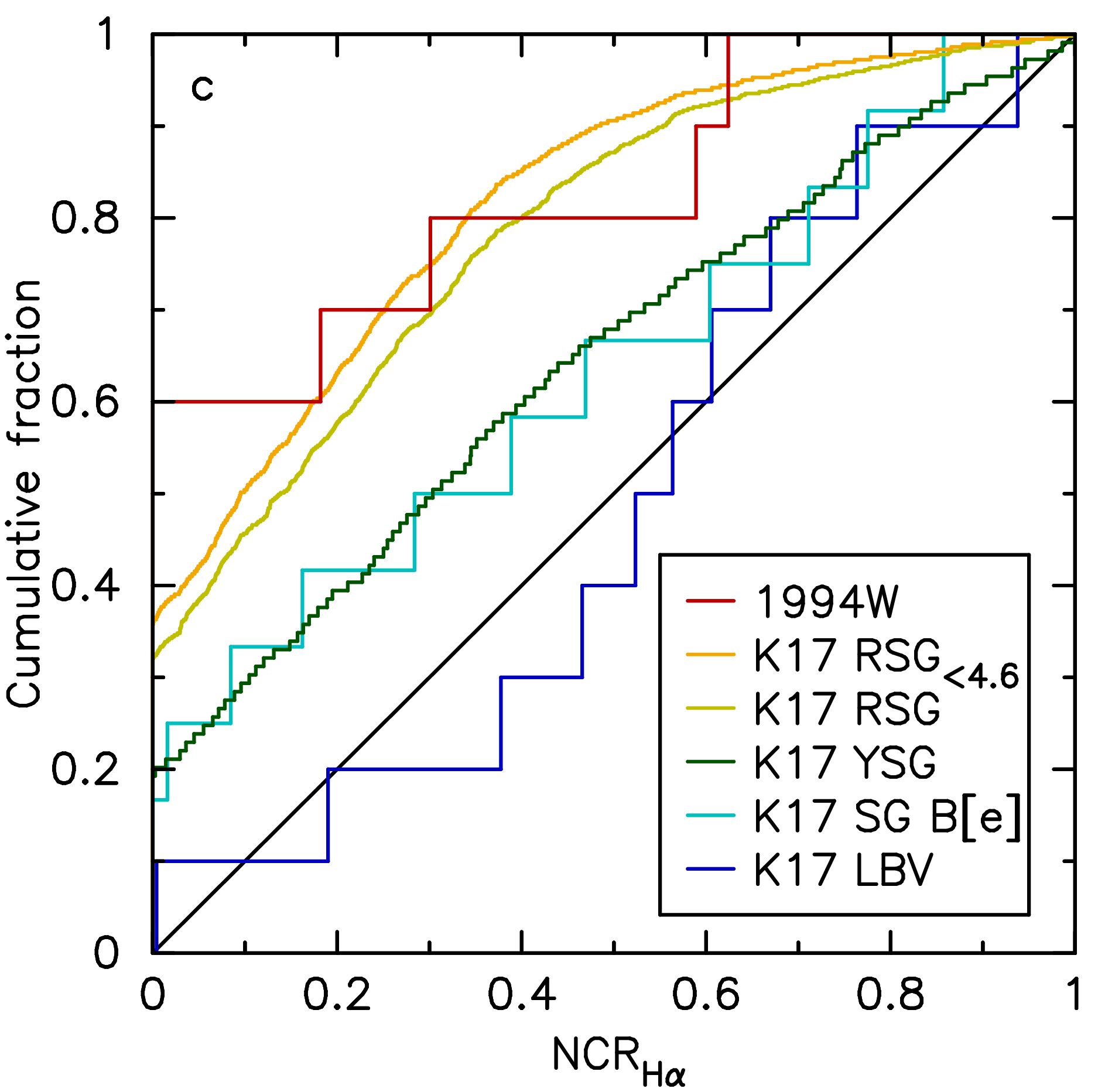}
\caption{NCR values of our sample of SN~1994W-like events. Comparisons include other main CCSN subtypes in a) NCR$_{\mathrm{H}\alpha}$ and b) NCR$_{\mathrm{NUV}}$ \citep[][indicated in the legend as A12, H14, and R22, respectively]{anderson12, habergham14, ransome22}. In c) NCR$_{\mathrm{H}\alpha}$ values of the SN~1994W-like sample are compared to those of a selection of massive stars in the LMC simulated at the distance of 60.1~Mpc based on the method and catalogues used by \citet[][indicated in the legend as K17]{kangas17}. For RSGs, both the complete sample and a subsample of less-massive RSGs with $\log(L/L_{\sun}) < 4.6$ are shown.}
\label{fig:ncr}
\end{figure*}

The resulting mean NCR$_{\mathrm{H}\alpha}$ value of our sample of SN~1994W-like events is $\langle$NCR$_{\mathrm{H}\alpha}\rangle = 0.170 \pm 0.076$, where the error is the standard error of the mean. The median distance of the sample galaxies is 60.1~Mpc, with a maximum distance of 76.8~Mpc. Our results can be compared to those of selected NCR studies with roughly similar median distances of Type IIP, Ib, and Ic SNe (at 21, 41, and 35~Mpc, respectively) by \citet{anderson12}, and Type IIn SN samples at 38~Mpc by \citet{habergham14} and at 62~Mpc by \citet{ransome22}. The main caveats in comparisons of different NCR values are related to the differences in the samples and the data. For example, if the sample distance increases, the spatial resolution decreases and the low or moderate NCR values are biased to increase while very high values may decrease. However, if the S/N ratio decreases, in particular the low NCR values are also biased to decrease. For a more detailed discussion of the caveats, see \citet{kangas17}. The cumulative NCR$_{\mathrm{H}\alpha}$ distribution of the sample of SN~1994W-like transients does not appear to trace the H$\alpha$ emission, and is shown in Fig.~\ref{fig:ncr} in comparison to reported values from the literature for a selection of CCSN subtypes. The $\langle$NCR$_{\mathrm{H}\alpha}\rangle$ value of the SN~1994W-like sample is in fact smaller than those reported for Type IIn \citep[$0.225 \pm 0.058$; $0.310 \pm 0.042$,][respectively]{habergham14,ransome22} or Type IIP \citep[$0.264 \pm 0.039$;][]{anderson12} SNe, marginally consistent within errors and subject to the distance effects. For comparison, Type Ic SNe, which are thought to arise from the massive end of the CCSN progenitor distribution, have a much stronger H$\alpha$ association \citep[$0.469 \pm 0.040$;][]{anderson12} that is inconsistent with that of our sample. 

\begin{table}
\caption{Our H$\alpha$ and NUV band NCR results of the SN~1994W-like sample with the JD values of the observations and the used ON and OFF H$\alpha$ filters indicated.}
\centering
\setlength\tabcolsep{2.5pt}
\begin{tabular}{lccccccc}
\hline
\hline
SN & JD$_{\mathrm{H}\alpha}$ & ON$_{\mathrm{H}\alpha}$ & OFF$_{\mathrm{H}\alpha}$ & JD$_{\mathrm{NUV}}$ & NCR$_{\mathrm{H}\alpha}$ & NCR$_{\mathrm{NUV}}$ \\
\hline
1994W & 2456245 & \#49 & \#52 & 2454241 & 0.182 & 0.554 \\ 
1994ak & 2456245 & \#49 & \#52 & 2453072 & 0.000 & 0.321 \\
1999eb & 2456244 & \#50 & \#21 & 2454407 & 0.589 & 0.875 \\
1999el & 2456244 & \#49 & \#52 & 2452875 & 0.000 & 0.440 \\
2003G & 2456244 & \#50 & \#21 & 2454390 & 0.000 & 0.000 \\
2004F & 2456244 & \#50 & \#21 & 2452972 & 0.624 & 0.647 \\
2004gd & 2456245 & \#50 & \#21 & 2453737 & 0.301 & 0.779 \\
2006bo & 2456244 & \#50 & \#21 & 2455816 & 0.000 & 0.297 \\
2009kn & 2456244 & \#50 & \#21 & - & 0.000 & - \\
2011ht & 2456245 & \#49 & \#52 & 2455304 & 0.000 & 0.478 \\
\hline
\end{tabular}
\label{table:ncr}
\end{table}

As an alternative approach, we carried out the NCR analysis for the sample using the Galaxy Evolution Explorer \citep[GALEX;][]{martin05} data in the near-UV (NUV) band (1750 to 2750~\AA) images (1$\farcs$5~pixel$^{-1}$). The GALEX pipeline reduced data were downloaded from the GALEX data release GR6/7 archive. GALEX data were not available for ESO~561-~G~020, the host galaxy of SN~2009kn. The NCR$_{\mathrm{NUV}}$ resulted in a mean value of $\langle$NCR$_{\mathrm{NUV}}\rangle = 0.488 \pm 0.084$. This is consistent with a flat distribution (mean NCR value of 0.5), similar to most CCSN subtypes with the exception of Type Ic SNe \citep[][see their figure 6]{habergham14}. The UV light probes quite recent (16 to 100~Myr) star formation \citep{gogarten09}. However, this is an older star formation phase than that traced by the H$\alpha$ light, and a flat NCR$_{\mathrm{NUV}}$ correlation for a SN subtype suggests that the bulk of their precursors have a relatively small progenitor mass. If rotating single stars in solar metallicity are assumed, the lifetime would suggest $M_\mathrm{ZAMS} \lesssim$14~$M_{\sun}$ progenitors \citep{ekstrom12}. However, due to the more complex evolution of binary systems, somewhat larger masses are also possible.

We compared the NCR$_{\mathrm{H}\alpha}$ values of SN~1994W-like events to those of a selection of different types of H-rich massive stars in the LMC. If the progenitor population of a SN type is drawn from among a particular stellar type, such stars relatively close to the end of their life cycle should have an NCR distribution consistent with that of the SNe. However, the NCR values are also influenced by factors unrelated to the stars or SNe themselves (e.g. pixel scale, spatial resolution, image depth, and SN position errors). Similar to \citet{kangas17}, we used the H$\alpha$ image of the LMC \citep{gaustad01}, which we simulated to have the typical resolution and image quality of our sample host galaxy data in order to have comparable NCR distributions. This was done by i) convolving the LMC image from its original spatial resolution of 58.2~pc to that obtained by ALFOSC at the median distance of 60.1~Mpc assuming a typical seeing of 1$\arcsec$ (289~pc); ii) rebinning the image from its original pixel scale of 11.6~pc~pixel$^{-1}$ to that of ALFOSC at this distance with a $2 \times 2$ binning (111~pc~pixel$^{-1}$); and iii) adding noise to the image to obtain a S/N ratio similar to our ALFOSC observations in typical star-forming regions. Catalogues of different massive stars in the LMC were used, including RSGs and yellow super giants \citep[YSGs;][]{neugent12}, B[e] supergiant (SG B[e]) stars \citep{zickgraf06,bonanos09} and LBVs \citep{smith15}, to calculate their NCR distributions from the simulated image, adding a Gaussian positional uncertainty of 0$\farcs$5 to their coordinates. We note the possible caveat that for example some YSGs or LBVs could be missing from the sample due to being obscured by dusty environments, which can have some effect on the NCR values. The RSGs were split into two subsamples based on their bolometric luminosity to carry out additional comparisons.

The resulting mean NCR$_{\mathrm{H}\alpha}$ values from the LMC data were $0.181 \pm 0.012$, $0.211 \pm 0.011$, $0.271 \pm 0.020$, $0.355 \pm 0.030$, $0.363 \pm 0.089$, and $0.510 \pm 0.083$ for the RSG$_{\log(L/L_{\sun}) < 4.6}$ ($M_\mathrm{ZAMS} \lesssim$9~$M_{\sun}$ stars), RSG, RSG$_{\log(L/L_{\sun}) \geq 4.6}$, YSG, SG~B[e], and LBV samples, respectively. We carried out an Anderson-Darling (AD) statistical test to compare the NCR$_{\mathrm{H}\alpha}$ distribution of the SN~1994W-like sample to that of the selected massive star. The resulting $p$-values from the AD test are 27 (RSG$_{\log(L/L_{\sun}) < 4.6}$), 18 (RSG), 6 (RSG$_{\log(L/L_{\sun}) \geq 4.6}$), 7 (SG~B[e]), 2 (YSG), and 1~\% (LBV). Therefore, the H$\alpha$ emission at the explosion sites of the SN~1994W-like sample is inconsistent with more massive stars, such as YSGs or LBVs (Fig.~\ref{fig:ncr}). However, the sample overlaps with the distribution of the RSGs. The NCR results also suggest that fallback explosions of relatively massive stars can be ruled out as a channel for the SN~1994W-like events.

\section{Spectrophotometric evolution of SN 1999eb}
\label{sect:99eb}

\begin{figure*}
\centering
\includegraphics[width=0.49\linewidth]{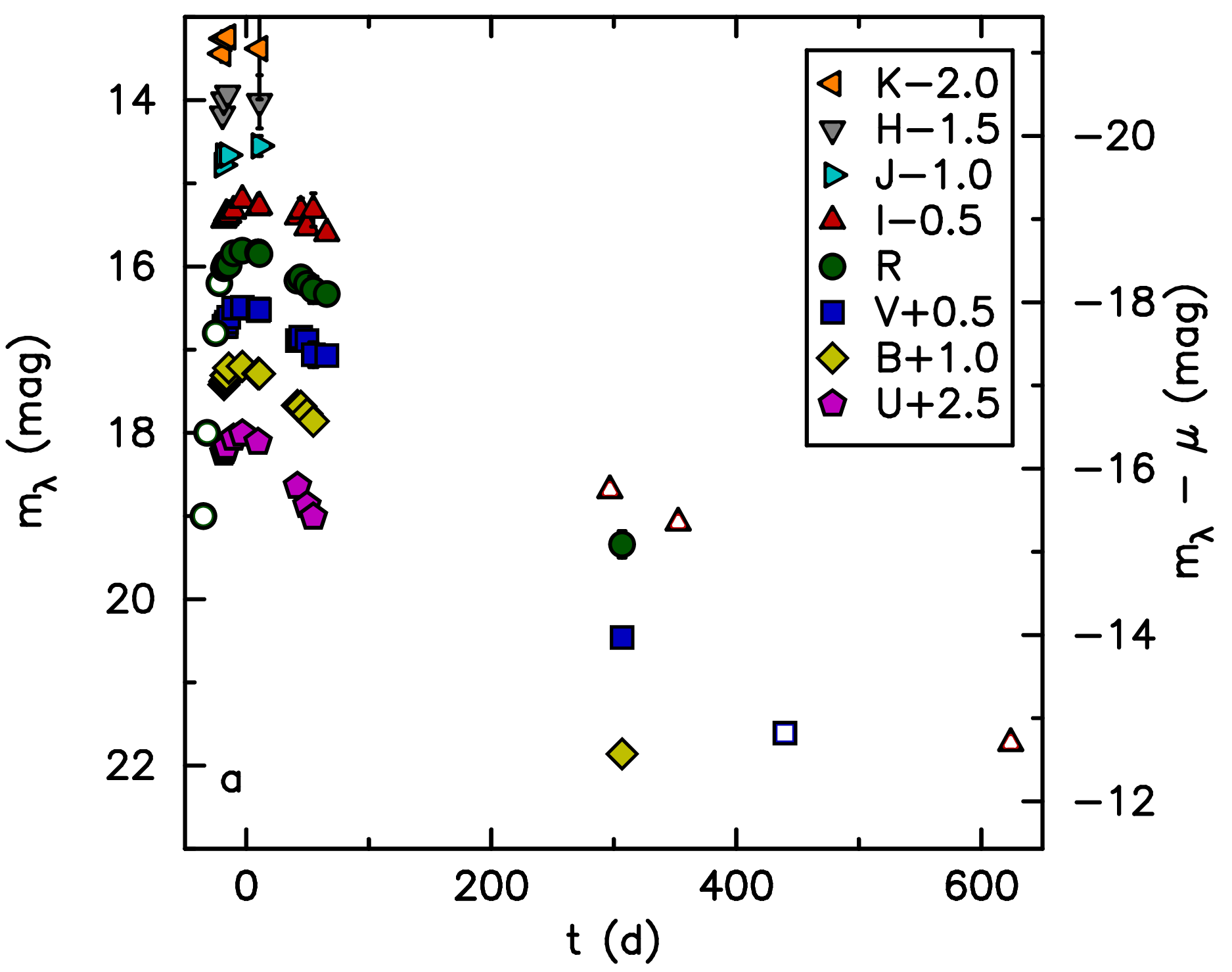}
\includegraphics[width=0.49\linewidth]{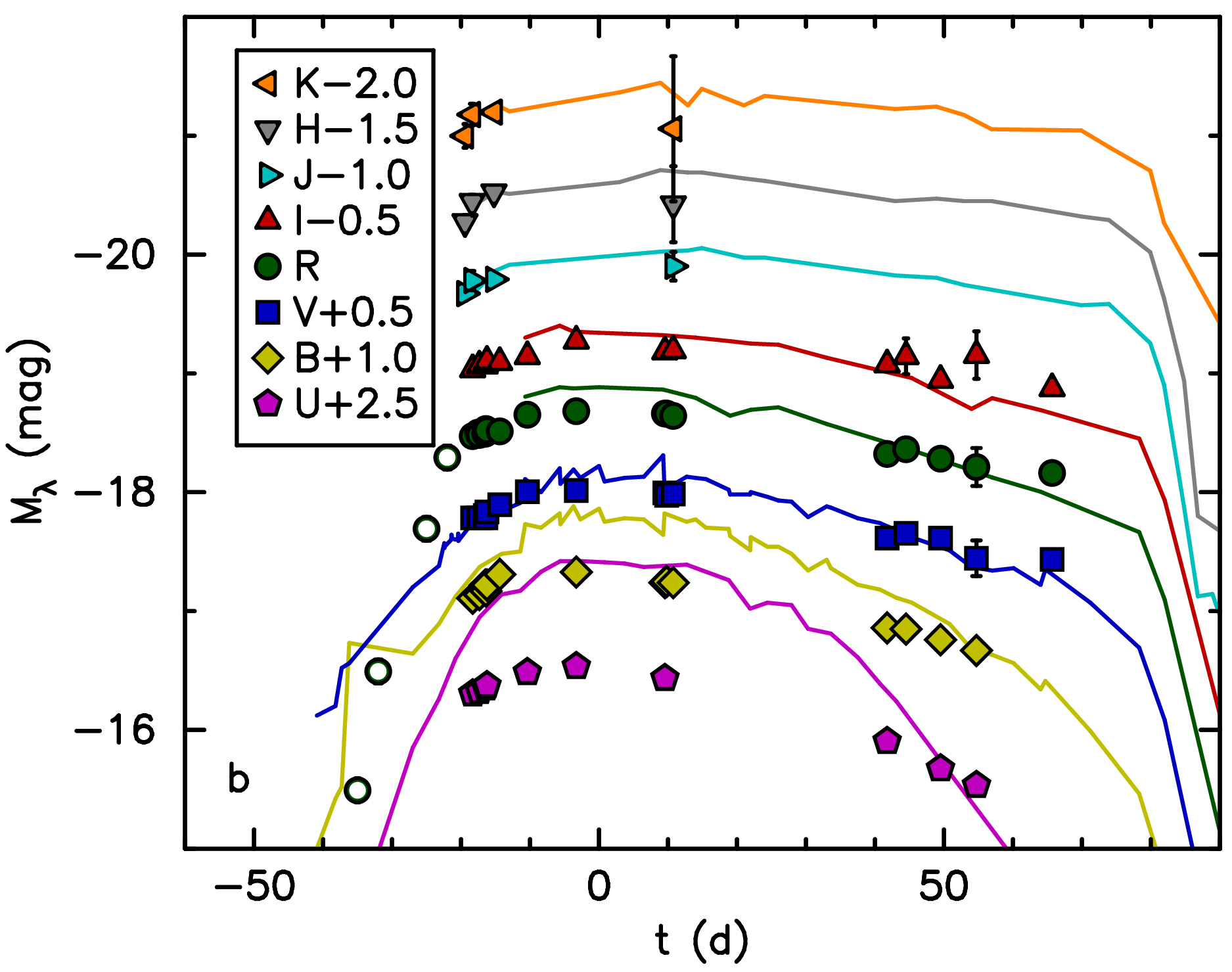}
\caption{a) Our observed light curves of SN~1999eb (solid symbols). For completeness, the early KAIT observations by \citet{modjaz99} and the late-time HST detections reported by \citet{li02} are also shown (open symbols). b) Absolute light curve evolution at the plateau of SN~1999eb (points) with the estimated negligible host galaxy extinction adopted. The comparison light curves (solid lines) of SN~2011ht \citep{roming12,mauerhan13} are shown for comparison and shifted vertically by $-1.6$~mag.}
\label{fig:SN1999eb}
\end{figure*}

\subsection{SN 1999eb data}

\begin{figure}
\centering
\includegraphics[width=\linewidth]{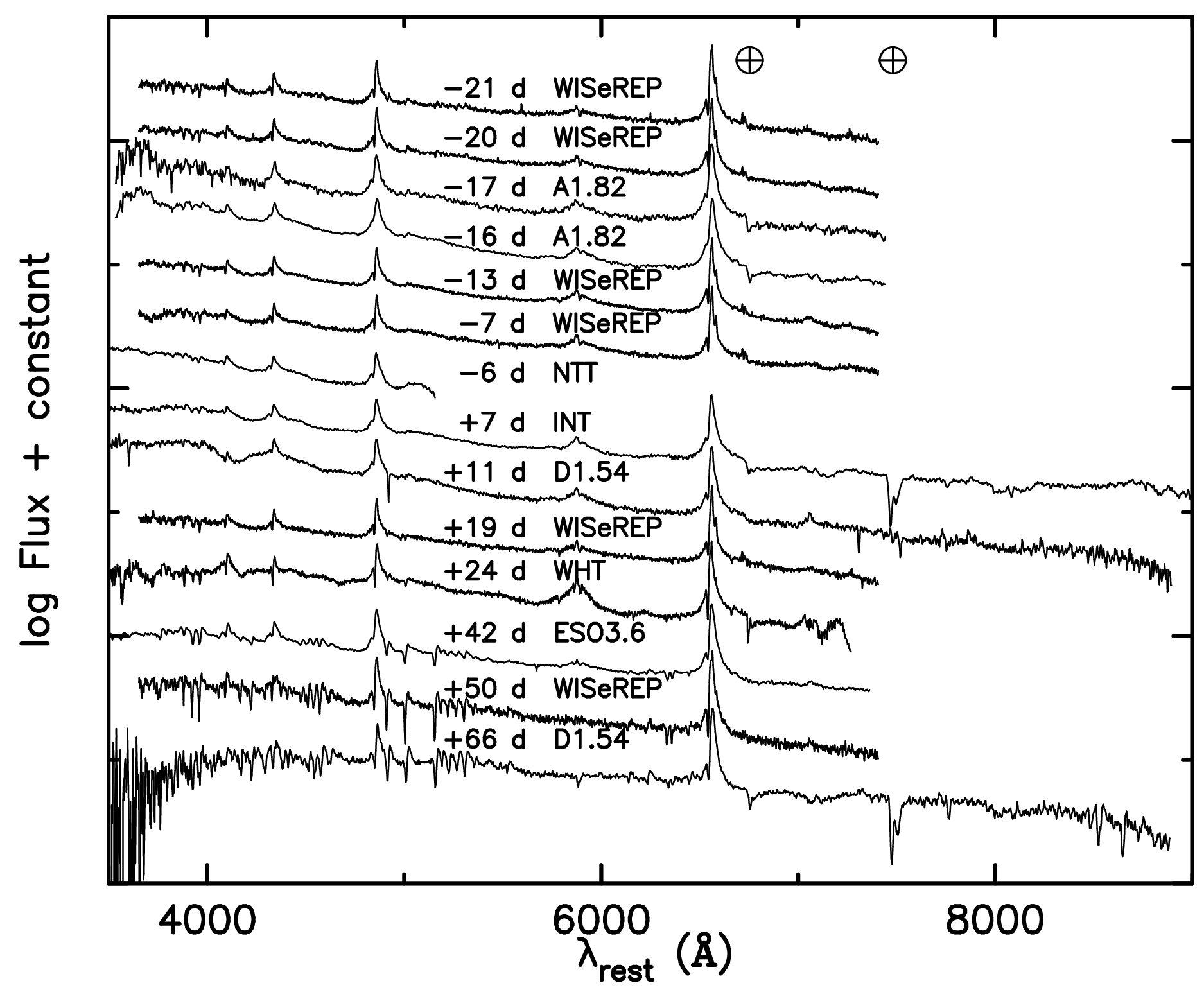}
\caption{Spectroscopic time series of SN~1999eb with epochs and used telescopes indicated. For completeness, the sequence includes also the public spectra on the WISeREP database (no available details on the used instruments). The spectra were dereddened by the Galactic extinction and the wavelengths corrected to the host galaxy rest frame. The wavelength of the most prominent telluric bands are indicated with a $\oplus$ symbol. For clarity, some of the lower quality spectra were omitted from the series and the spectra have been vertically shifted.}
\label{fig:spect_99eb}
\end{figure}

The photometric data of SN~1999eb is reported in Table~\ref{table:phot_99eb} and shown in Fig.~\ref{fig:SN1999eb}. The observations were obtained using the 3.6~m Telescopio Nazionale Galileo (TNG) with the Optical Imager Galileo (OIG) and the ARcetri Near Infrared CAmera (ARNICA), the Asiago 1.82~m Copernico Telescope with the Asiago Faint Object Spectrograph and Camera (AFOSC), the European Southern Observatory (ESO) 3.6~m telescope with the ESO Faint Object Spectrograph and Camera 2 (EFOSC2), the 8~m Very Large Telescope (VLT) with the Infrared Spectrometer And Array Camera \citep[ISAAC;][]{moorwood98b}, the Danish 1.54~m telescope with the Danish Faint Object Spectrograph and Camera (DFOSC), and the 3.58~m New Technology Telescope (NTT) with the Son of ISAAC \citep[SOFI;][]{moorwood98a}. Our spectroscopic log of SN~1999eb observations is reported in Table~\ref{table:spect_99eb} and the spectral time series is shown in Fig.~\ref{fig:spect_99eb}. The data was obtained with the Asiago 1.82~m telescope with AFOSC, the ESO 3.6~m with EFOSC2, the NTT with the ESO Multi-Mode Instrument (EMMI), the 2.54~m Isaac Newton Telescope (INT) with the Intermediate Dispersion Spectrograph (IDS), and the 4.2~m William Herschel Telescope (WHT) with the Intermediate dispersion Spectrograph and Imaging System (ISIS).

Standard imaging and spectroscopic data reductions of SN~1999eb were carried out using {\sc iraf} tasks. The images were calibrated using the optical \textit{ugri} magnitudes of a selection of field stars from the Sloan Digital Sky Survey \citep[SDSS;][]{eisenstein11} Data Release 12 \citep[DR12;][]{alam15}, and near-IR \textit{JHK} magnitudes from the Two Micron All Sky Survey \citep[2MASS;][]{skrutskie06}. The SDSS magnitudes were converted into the Johnson-Cousins \textit{UBVRI} system using the transformations of \citet{jester05}. Archival \textit{BVRI} template images of the field were made use of; images were obtained using the ESO 3.6~m telescope with EFOSC2 on 2002 September 13.23~{\sc ut}. The host NGC~664 is less luminous and more compact at shorter \textit{U}-band and longer \textit{JHK}-band wavelengths, and template subtraction was not a necessary process to carry out in these bands. The {\sc isis} 2.2 package was used to carry out the image subtractions. The PSF photometry was carried out using a slightly modified version of the {\sc quba} pipeline \citep{valenti11}. The spectra were absolute flux calibrated based on broad-band photometry, and corrected for Galactic extinction and host redshift.

\subsection{SN 1999eb results}

The photometry of SN~1999eb reveals a flat plateau over a time range of roughly 85~d. SN~2011ht shows a similar light curve evolution to that of SN~1999eb, although the former appears to show an intrinsic blue excess at the shortest wavelengths \citep{mauerhan13}. We estimated the line-of-sight host galaxy reddening of SN~1999eb with a simultaneous $\chi^{2}$ fit of the \textit{VRIJHK} light curves to those of SN~2011ht. The method has three free parameters: the host galaxy extinction, $A_{V,\mathrm{host}}$; the intrinsic magnitude difference of the two SNe applied to all the compared bands, $C$; and the time, $t$, at the discovery in comparison to a chosen reference point, $t_{0}$. The 1$\sigma$ errors were estimated from the $\chi^{2}$ fit based probability density functions of the free parameters. For further details of the method, see \citet{kankare14a,kankare14b}. The comparison suggests an approximate light curve maximum for SN~1999eb at JD~2451475.9, a negligible host galaxy extinction of $A_{V} = 0.0^{+0.2}_{-0.0}$~mag, and that SN~1999eb was $1.6^{+0.2}_{-0.1}$~mag brighter than SN~2011ht. The resulting plateau maximum for SN~1999eb is $M_{R} \approx -18.7$~mag. The absolute light curves of SN~1999eb at the plateau are shown in Fig.~\ref{fig:SN1999eb} with shifted light curves of SN~2011ht shown in comparison. 

\begin{figure}
\centering
\includegraphics[width=\linewidth]{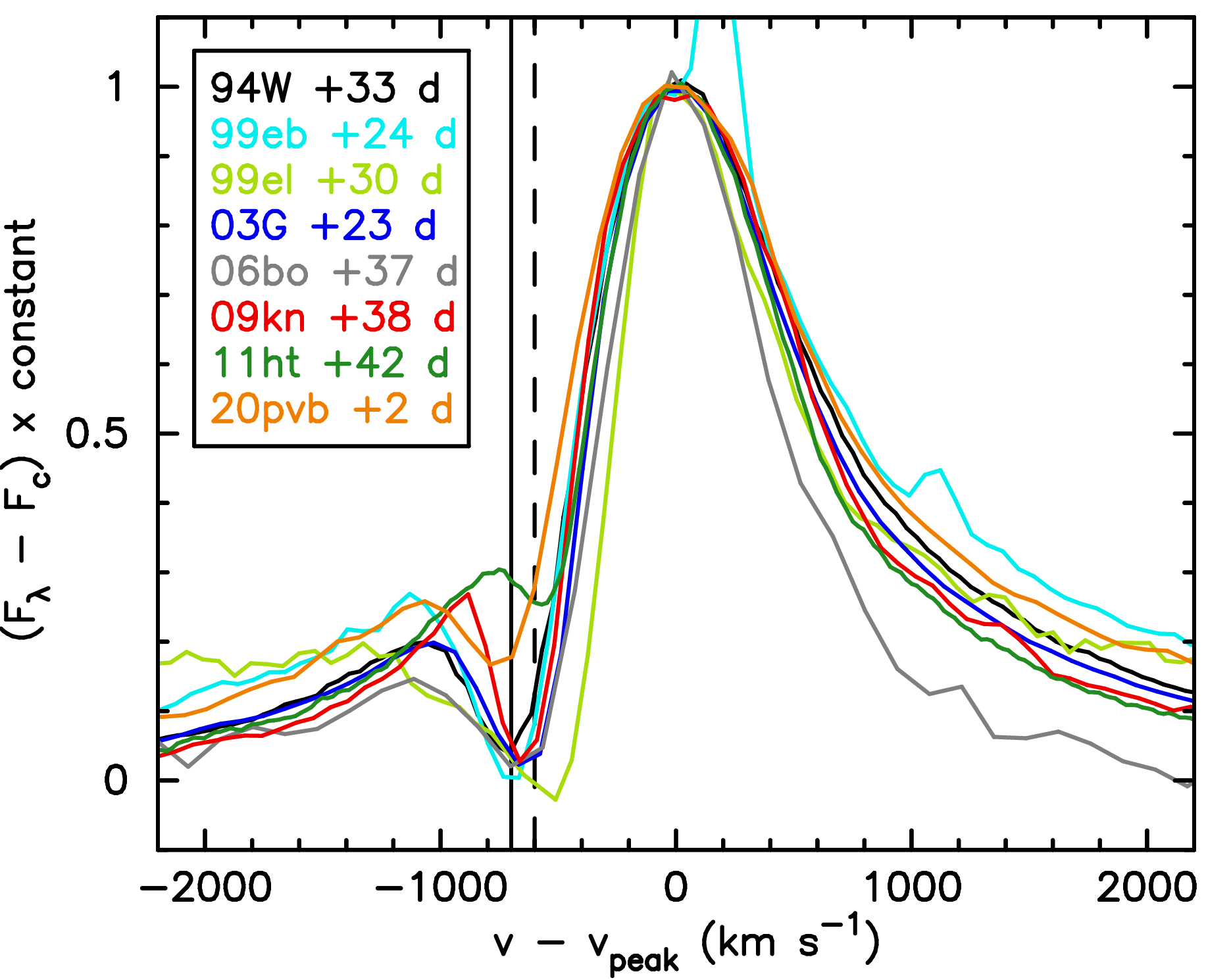}
\caption{Comparison of continuum subtracted and peak normalised H$\alpha$ profiles of SN~1994W-like events primarily around +35~d with a spectral resolution of $R \geq 800$. The H$\alpha$ lines show similarly blueshifted P~Cygni absorption minima around 600 to 700~km~s$^{-1}$ (dashed and solid vertical lines, respectively). The spectral profile of SN~1999eb is contaminated by the host galaxy H$\alpha$ line and the effect of the feature was estimated and excluded from the normalization process.}
\label{fig:Halpha-comp}
\end{figure}

Non-terminal stellar outbursts are not expected to synthesize radioactive $^{56}$Ni, which decays into $^{56}$Co and subsequently into $^{56}$Fe \citep[e.g.][]{diehl98}, dominating the tail-phase light curve evolution with a theoretical bolometric luminosity decline rate of 0.98~mag~(100~d)$^{-1}$ assuming complete $\gamma$-ray trapping. The late-time \textit{I}-band decline rate of 0.94~$\pm$~0.01~mag~(100~d)$^{-1}$ for SN~1999eb is close to this value while the \textit{V}-band observations suggests a notably slower decline of 0.67~$\pm$~0.15~mag~(100~d)$^{-1}$. However, SN~1999eb is also quite luminous in the late-time observations (i.e. around $-$13.7, $-$14.5, and $-$15.2~mag in \textit{B}, \textit{V}, and \textit{R}-band, respectively, at +307~d from maximum light in the observer frame or +302~d in the rest frame). This suggests that some CSM interaction is likely still ongoing; unfortunately, to our knowledge, no spectroscopy of SN~1999eb during the tail-phase is available. We estimate a \textit{BVRI} pseudo-bolometric luminosity of the event using the {\sc SuperBol} package \citep{nicholl18} at the tail phase and assume the first epoch of detection as the explosion. We find a conservative estimate for the $^{56}$Ni upper limit of $<$0.3~$M_{\odot}$ for SN~1999eb by scaling the resulting luminosity, $L_{BVRI}$, to that of SN~1987A constructed from the corresponding late-time observations \citep{whitelock88} and assuming a $^{56}$Ni mass of 0.069 $M_{\odot}$ for SN~1987A \citep{bouchet91}. 

The typical spectral resolution of most of the observations of SN~1999eb was relatively low, within the range of $R = 200$ to $500$. The spectral evolution of SN~1999eb is similar to that of SN~1994W-like events, as seen in Fig.~\ref{fig:sample}. Close to light curve maximum, roughly around $-20$ to $+20$~d range, the spectra are dominated by strong Balmer lines with Lorentzian wings typical for electron scattering dominated emission line profiles. Furthermore, a narrow absorption component is present and has a velocity of roughly $-700$~km~s$^{-1}$ which is typical for SN~1994W-like events. Fig.~\ref{fig:Halpha-comp} shows a compilation of H$\alpha$ line profiles of SN~1994W-like events primarily close to $\sim$35~d from maximum light. The P~Cygni absorption component minimum velocities are typically within $\sim$600 to 700~km~s$^{-1}$ compared to the line maximum, which makes this narrow velocity range a distinct characteristic of these events. Higher-quality spectra were selected for the comparison (a spectral resolution of $R \geq 800$), since data with very low resolution will result in notable line broadening and overestimation of the velocity. For example, SNe~1994W, 2005cl, and 2011ht show a clear \HeI\ $\lambda$5875 feature, which is also the case with SN~1999eb through the spectral sequence up to +42~d. \FeII\ features can be visible during the epochs covered by the spectral time series; however, spectral resolution has an effect on how prominently these lines can be detected. At the post-peak plateau phase the metal P~Cygni lines become more prominent, dominated by \FeII\ and \CaII\ features, including the \CaII\ NIR triplet that is clearly visible in the +66~d spectrum of SN~1999eb, which is the only spectrum several tens of days past maximum that extends sufficiently far in red wavelengths to cover this line region.

\section{Early spectrophotometric nature of SN 2003G}
\label{sect:03G}

Our observations cover two epochs of imaging of SN~2003G using the Asiago 1.82~m telescope with AFOSC. We carried out standard data reductions and PSF photometry using the {\sc quba} pipeline. The photometry was calibrated using the \textit{ugri} field stars magnitudes from the SDSS DR12, converted into the \textit{BVRI} system \citep{jester05}. This resulted in the magnitudes on JD~2452664.36 of $m_{B} = 15.27 \pm 0.01$, $m_{V} = 15.15 \pm 0.04$, $m_{R} = 14.92 \pm 0.04$, and $m_{I} = 14.81 \pm 0.02$~mag, and on JD~2452696.27 of $m_{B} = 15.42 \pm 0.04$, $m_{V} = 15.26 \pm 0.03$, $m_{R} = 15.03 \pm 0.02$, $m_{I} = 14.97 \pm 0.04$~mag. The light curve reveals a rapid rise followed by a flattening, very similar to that of SN~1999eb. We carried out a simultaneous $\chi^{2}$ fit of the SN~2003G \textit{BVRI} light curves to those of SN~1999eb, similar to the process between the latter and SN~2011ht (Sect.~\ref{sect:99eb}). This suggested that SN~2003G had a negligible host extinction of $A_{V} = 0.0^{+0.1}_{-0.0}$~mag, reached the light curve maximum around JD~2452675.7, and was $0.1 \pm 0.1$~mag brighter than SN~1999eb (Fig.~\ref{fig:03G}).\footnote{For completeness, we carried out a similar analysis to the two epochs of AFOSC imaging of SN 2004gd, which resulted in the magnitudes on JD~2453355.04 of $m_{B} = 18.64 \pm 0.07$, $m_{V} = 17.94 \pm 0.04$, $m_{R} = 17.50 \pm 0.04$, and $m_{I} = 17.24 \pm 0.04$~mag, and on JD~2453373.96 of $m_{B} = 18.96 \pm 0.05$, $m_{V} = 18.11 \pm 0.04$, $m_{R} = 17.70 \pm 0.04$, $m_{I} = 17.37 \pm 0.05$~mag. The light curve comparison with SN~1999eb suggested that SN~2004gd had a host extinction of $A_{V} = 1.0 \pm 0.2$~mag, peaked roughly around JD~2453324.5, and was $0.6^{+0.2}_{-0.1}$~mag fainter than SN~1999eb (Fig.~\ref{fig:03G}).} 

\begin{figure*}
\centering
\includegraphics[width=\linewidth]{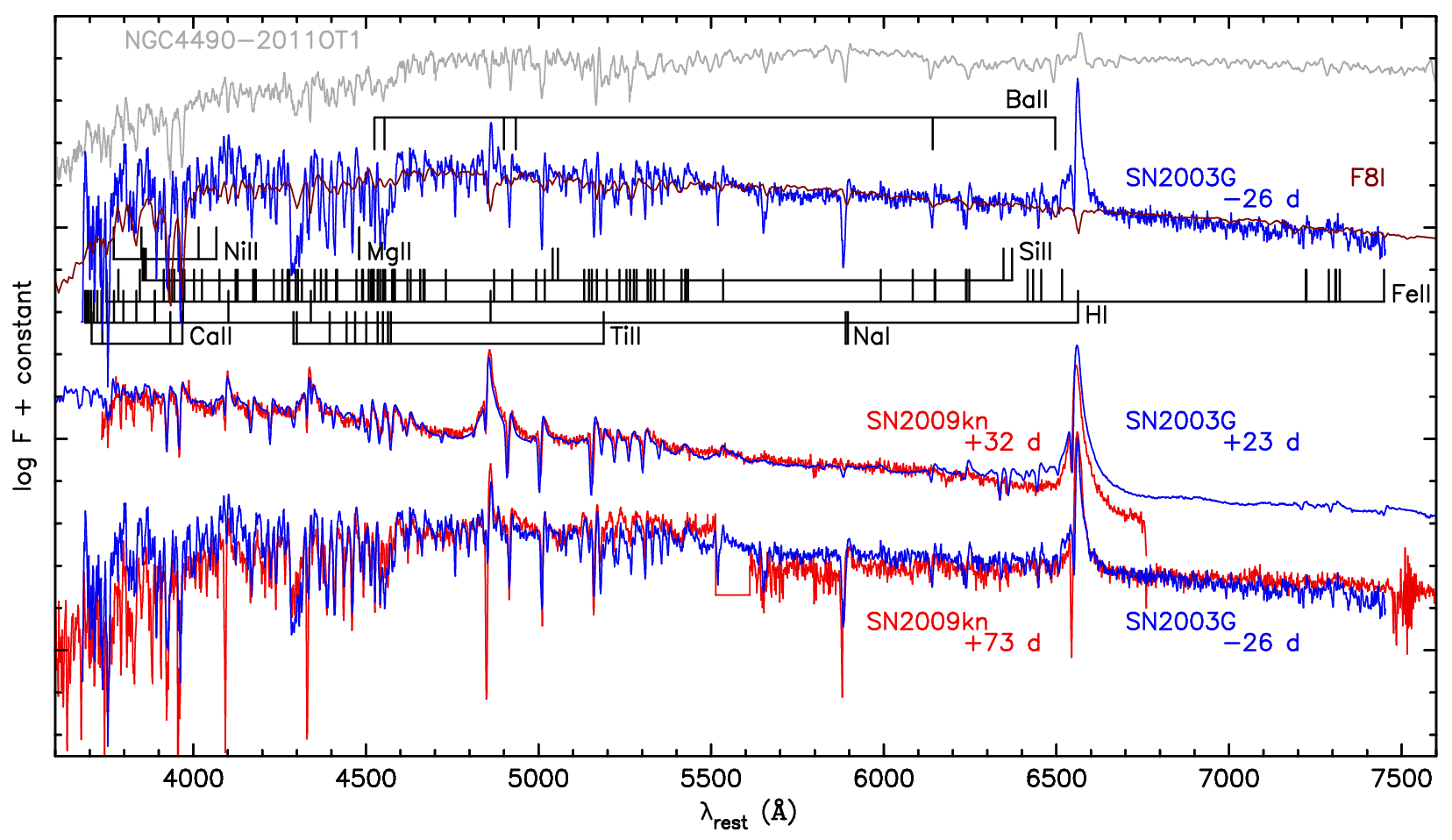}
\caption{Early $-26$~d spectrum of SN~2003G compared to LRN NGC~4490-2011OT1 close to the red maximum \citep{pastorello19b}, and to a F8~I type supergiant template \citep{pickles98}. SN~2003G evolves spectroscopically into a SN~1994W-like event as shown by the comparison to SN~2009kn \citep{kankare12}. The early $-26$~d spectrum of SN~2003G also shows similarities to SN~2009kn close to the end of the plateau at +73~d. A line list used by \citet{dessart09} to model a synthetic late-plateau spectra of SN~1994W is shown as a reference for the early spectrum of SN~2003G, in addition to our identification of the plausible \BaII\ features.}
\label{fig:SN2003G}
\end{figure*}

Four epochs of archival spectra of SN~2003G are available via the WISeREP database. Based on the light curve evolution of the event, the spectra have been obtained at epochs $-26$, $-8$, $-1$, and $+23$~d from the light curve maximum, with the latter three reported by \citet{shivvers17} in a SN sample study; these observations confirm the SN~1994W-like spectroscopic nature of the transient, as shown in Fig.~\ref{fig:sample}. The $-26$~d spectrum of SN~2003G is distinctively redder than the spectra near maximum. A large number of P~Cygni lines are visible in the $-26$~d spectrum with absorption minimum velocities of $\sim$500~km~s$^{-1}$. Identifiable features consist of H$\alpha$, H$\beta$, \NaID, \CaII\ H\&K, and a plethora of other metal lines, including multiplets of \FeII, \TiII, and \SiII. We also suggest that the spectrum includes a few \BaII\ features. A P~Cygni feature consistent with \BaII\ $\lambda$6497 is distinctly separated from the \FeII\ $\lambda\lambda$6456,6516 lines. \BaII\ $\lambda$6142 is blended with \FeII\ $\lambda\lambda$6148,6149, and \BaII\ $\lambda$4934 could be blended with \FeII\ $\lambda$4924. The \BaII\ $\lambda$4900 line can be identified, separate from \FeII\ $\lambda$4924. The \BaII\ lines $\lambda$4525, $\lambda$4554, and $\lambda$4934 may be blended with a selection of nearby \FeII\ and \TiII\ features. \BaII\ $\lambda$5854 is not prominent; however, it can be weak compared to many aforementioned \BaII\ features \citep{davidson92}. \BaII\ lines are commonly seen in LRNe \citep{blagorodnova17,blagorodnova21,cai19,cai22,pastorello19b,pastorello21a,pastorello21b,pastorello23}, but also in low-luminosity Type IIP SNe \citep[e.g.][]{valerin22} and in SN~1987A-like SNe \citep[e.g.][]{sit23}. \BaII\ lines in peculiar Type II SNe have been associated with relatively low effective temperatures \citep[e.g.][]{turatto98} and dense, low-velocity ejecta, potentially enhanced by clumping \citep{dessart19}. The $-26$~d spectrum of SN~2003G shows a prominent and broad absorption feature around 4300~\AA, which is seen in many SN~1994W-like events at late-plateau phases, and is likely a blend of \FeII\ and \TiII\ features \citep{dessart09}.

\citet{pastorello19b} noted that the very early spectrum of SN~2011ht, for which we adopted an epoch of $-45$~d, closely resembled the LRN NGC 4490-2011OT1 near the red, second light curve peak. Such an overall similarity is also shown by the early spectrum of SN~2003G, though the spectrum is not quite as red and in particular the H$\alpha$ line is more prominent; however, the spectrum is not obtained at as early an epoch as the first spectral observations of SN~2011ht. Nonetheless, this also shows that the early spectral characteristics of SN~2011ht are not unique among SN~1994W-like events. The $-26$~d spectrum of SN~2003G also shows similarities to those of SN~1994W-like events close to the end of the plateau phase, see Fig.~\ref{fig:SN2003G}. \citet{elias-rosa24} reported the same effect for the similar $-21$ and +31~d spectra of SN~2020pvb. However, compared to SN~2003G, the rise time of $\sim$60~d to the maximum light of SN~2020pvb is relatively long and the spectral evolution of that event is slower. 

\citet{humphreys12} noted the similarity of the spectrum of SN~2011ht at the end of the plateau phase to those of late F-type and early G-type supergiants. Motivated by the similarity of the $-26$~d spectrum of SN~2003G to those of the late-plateau spectra of SN~1994W-like events, we scaled the early spectrum to the template spectra of F5~I, F8~I, G0~I, G2~I, and G5~I supergiants from \citet{pickles98}, and found the best match with the F8~I template; the distinctive differences are the more prominent metal line features in the SN~2003G spectrum and, in particular, the strong P~Cygni Balmer lines. In Fig.~\ref{fig:SN2003G}, the spectra of SN~2003G are compared to those of LRN NGC~4490-2011OT1, type F8~I supergiant template, and SN~2009kn.

By the time of the second $-8$~d spectrum, SN~2003G has evolved to be bluer and the metal lines have largely disappeared with the exception of the most prominent \FeII\ lines of the $\lambda\lambda$4924,5018,5169 multiplet. Furthermore, the Balmer line series has become more prominent, with less prominent absorption components compared to the overall line profiles. No major evolution is evident between the $-8$ and $-1$~d spectra (Fig.~\ref{fig:sample}). The +23~d spectrum resembles SN~1994W-like events at early plateau, as indicated by the strong similarity to the +32~d spectrum of SN~2009kn, see Fig.~\ref{fig:SN2003G}.

\section{Explosion sites of SN 1994W-like transients in HST data}
\label{sect:HST}

The Mikulski Archive for Space Telescopes was searched for high-resolution HST observations which covered the explosion site of SN~1994W-like transients. The aim was to use images to identify associated sources or set non-detection limits. The analysed data was obtained with the Wide Field and Planetary Camera 2 (WFPC2), the Advanced Camera for Surveys (ACS), or the Wide Field Camera 3 (WFC3). 

\subsection{Late-time HST imaging of SN 1994W}
\label{sect:94W}

There are post-explosion HST images of the site of SN~1994W obtained with the WFPC2 in F450W, F606W, and F814W bands on 2001 July 4.8~{\sc ut}, the ACS in F814W band on 2004 May 29.6~{\sc ut}, and the WFC3 in F336W band on 2011 January 30.0~{\sc ut}. For the first measurement of the position of SN~1994W, we used imaging of SN~1994W from the Jacobus Kapteyn Telescope (JKT) with the TEK4 instrument on 1994 November 16. The 0$\farcs$33 pixel$^{-1}$ scale and $\sim$6\arcmin$\times$6\arcmin\ FOV of the JKT image are much larger than those of the HST data; therefore, an intermediate transformation was used to a deep \textit{R}-band image of NGC~4401 from the NOT with ALFOSC on 2009 January 14 (kindly provided by J. Knapen), before transforming the event location to the drizzled WFPC2 F606W mosaic. To measure the transient position in the JKT image, the local background was first fitted and subtracted with a two-dimensional polynomial. The pixel coordinates of the location were measured using three different centering algorithms in the {\sc digiphot} package in {\sc iraf}, taking the average value as the event position and the standard deviation of the three values (78~mas) as the uncertainty. To identity the position of SN~1994W in the HST images, we used sources common to the JKT and the NOT, and the NOT and the HST images to derive a geometric transformation with the {\sc geomap} task in {\sc iraf}, allowing for shifts, rotation, and a scale factor. 14 sources common to the JKT and the NOT images were used, giving an uncertainty of 0$\farcs$26 for the transformation; 23 sources were identified in both the NOT and WFPC2 images, giving a 0$\farcs$10 error. The total uncertainty for the position of SN~1994W in the WFPC2 F606W image is hence 0$\farcs$29.

\begin{figure*}[!h]
\centering
\includegraphics[width=\linewidth]{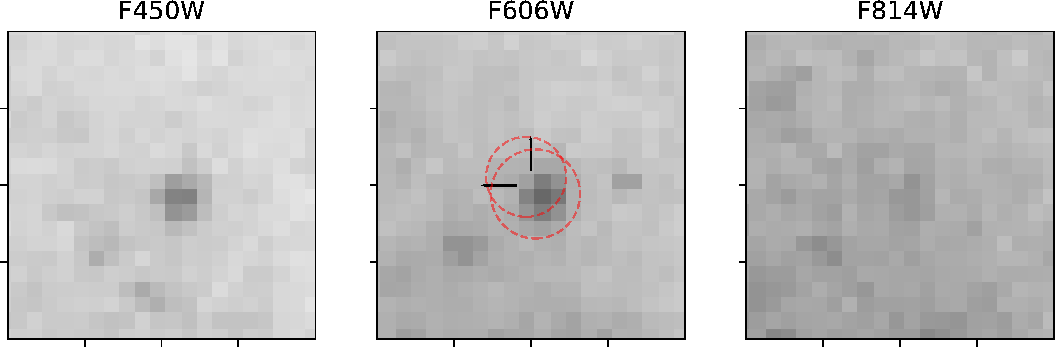}\\
\vspace{+0.5cm}
\includegraphics[width=\linewidth]{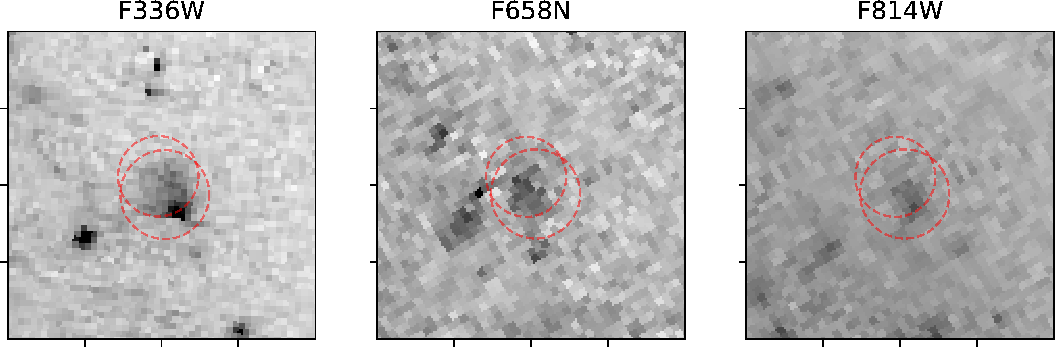}
\caption{The top row shows WFPC2 cutouts and the bottom row the ACS and WFC3 cutouts centred on the position of SN~1994W with a size of $2 \times 2$~arcsec$^{2}$, and an orientation of north up and east left. The candidate host cluster of SN~1994W is most clearly seen in the F336W filter, with the point-like source to the southwest, and the extended diffuse emission to the north-east. In all panels the two red dashed circles indicated the 1$\sigma$ uncertainty on the transformed position of SN~1994W, as determined from the JKT and NOT images.}
\label{fig:hst_94W}
\end{figure*}

The second measurement of the position of SN~1994W was obtained from a late time image from the NOT with STANCAM1 on 1995 January 27. At this late epoch, the event was quite faint, and the uncertainty in its location is much larger. However, we were able to align the NOT image directly to the HST data, thus eliminating the need for an intermediate transformation. To measure the position of the SN in the NOT image, the ALFOSC \textit{R}-band image was convolved and scaled to match the seeing of the STANCAM1 image, and subtracted from the latter. The centroid of SN~1994W was measured in the subtracted frame using three different algorithms as before, and transformed to the HST frame. The standard deviation of the three measurements of the position was 58~mas, while the error in the transformation was 0$\farcs$25. Hence, the total uncertainty for the location of SN~1994W is 0$\farcs$26.

In both cases, the position in the WFPC2 data is consistent with a relatively bright $\sim$22~mag point-like source in the field (Fig.~\ref{fig:hst_94W}). {\sc dolphot} was used to fit a PSF to this source and the measured magnitudes are listed in Table~\ref{table:hst}. To determine the position of SN~1994W in the ACS and WFC3 images, a transformation was determined between these images and the drizzled WFPC2 F606W mosaic. The uncertainty in this transformation was $\sim$10 to 20~mas, and consequently negligible compared to the uncertainty due to the ground-based data. The superior resolution of the ACS and WFC3 data shows that the source seen in WFPC2 is comprised of a point-like source and some resolved, extended emission (Fig.~\ref{fig:hst_94W}). {\sc dolphot} finds a best fit to the emission complex in the ACS data using three point sources. Interestingly, the combination of the three F814W magnitudes ($24.02 \pm 0.12$, $24.73 \pm 0.21$, and $25.18 \pm 0.32$~mag) measured in the ACS image results in total $m_{\mathrm{F814W}} = 23.34 \pm 0.11$~mag, which is consistent with the photometry of the WFPC2 image, and suggest that the source is likely a cluster or some other complex structure. The absolute magnitude of $M_{\mathrm{F814W}} = -8.6$~mag is also consistent with what one expects for a cluster. In the WFC3 F336W image, {\sc dolphot} fits the point source to the southwest with a PSF, but also fits the extended emission with 11 fainter point sources. Examination of the residual image after PSF subtraction reveals that the combination of these faint sources cleanly model the diffuse flux. The combined magnitudes of these faint sources are reported in Table~\ref{table:hst}.

\begin{figure*}
\centering
\includegraphics[angle=90,width=\linewidth]{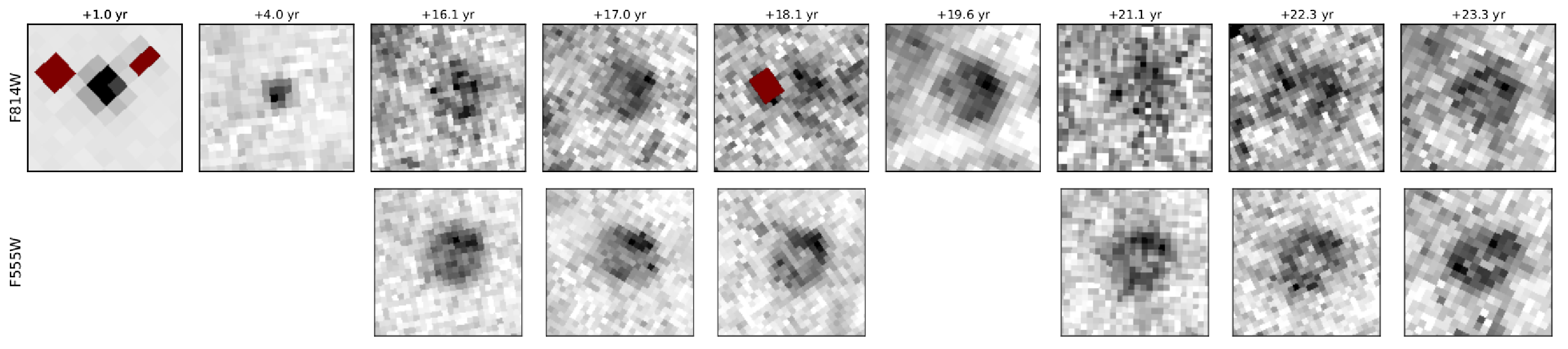}
\caption{Panels show a $1 \times 1$~arcsec$^{2}$ cutout in F555W and F814W filters, centered on the location of SN~1999el with north to the left and east down. Cosmic rays are masked in maroon, the phase of each image with respect to the discovery epoch is indicated above each column.}
\label{fig:hst_99el}
\end{figure*}

To further explore the nature of the possible cluster close to SN~1994W, we used the colour-magnitude diagram (CMD) interface to the PARSEC stellar evolutionary models \citep{bressan12}. The expected magnitudes in the HST filters were calculated for a cluster at solar metallicity that formed during a single burst of star formation, and the spectral energy distributions (SEDs) were scaled to match the F606W magnitude (Fig.~\ref{fig:cluster_94W}). For population ages between 6.0 and 6.6~dex, and scaling the cluster mass to a value between 10000 and 4300~$M_{\sun}$ (where a larger scaling is needed for a younger cluster), we can match the broad-band photometry with the exception of the F336W band, which is over-predicted by the model by 1.0 to 1.5~mag. We disfavour an older population ($\log(\mathrm{Age/yr})>6.8$~dex) as these models have redder colours at optical wavelengths. Compared to the model SEDs, the narrow-band F658N flux excess arise likely from the H$\alpha$ line emission.

The \textit{Gaia} space observatory Data Release 3 \citep[DR3;][]{gaia23} reported over 40 sources in the FOV of the WFC3 F336W image of the field of SN~1994W. Based on these sources the world coordinate system (WCS) of the calibrated HST archival frame appears to be quite robust, and based on our derived pixel coordinates of the location of the transient our best estimate for the revised coordinates of SN~1994W are $\alpha = 12^{\mathrm{h}}02^{\mathrm{m}}11\fs01$ and $\delta = +62\degr08\arcmin31\farcs8$, which we adopted for our NCR analysis (Sect.~\ref{sect:ncr}).

\subsection{Late-time HST imaging of SN 1999el}

SN~1999el appears as a point source in the first two epochs of HST imaging at +1.0 and +4.0~yr (Fig.~\ref{fig:hst_99el}). The field was revisited by HST at +16.1~yr and a striking light echo had developed. The echo appears broadly circular, with slightly more emission to the southwest. The diameter of the echo remains fairly constant at $\sim$0$\farcs$4 over the seven years from +16.1 to +23.3~yr. There is no sign of the light echo in the WFC3 broad-band F110W and narrow-band F128N observations taken on 2016 March 5.2~{\sc ut}. Using the standard formulae to describe the radius evolution of a light echo \citep[e.g.][]{vanloon04}, we find that what we observe is broadly consistent with scattering by either a thin sheet or shell of dust at a radial distance of between 50 and 100~pc from SN~1999el. While SN~1999el is among the closest events in our sample, it is not possible to constrain further the properties and geometry of the scattering material; unlike in some other examples \citep[e.g.][]{stritzinger22} multiple echoes are not resolved. However, the light echo is qualitatively consistent with the high host extinction inferred towards this event. The light echo is an extended source; therefore, photometry with {\sc dolphot} is not feasible. However, we measured with {\sc dolphot} magnitudes of a selection of isolated, nearby point-like sources within a similar magnitude range to that of the light echo. We proceeded to measure these sequence sources and the light echo with aperture photometry using the Graphical Astronomy and Image Analysis ({\sc gaia}) tool with an aperture diameter of $\sim$0$\farcs$6. Resulting magnitudes from this are reported in Table~\ref{table:hst}. Recently, \citet{baer-way24} reported HST observations of a sample of SNe and noted that SN~1999el was not detected in the F555W and F814W band observations of SN~2020dpw in NGC~6951 on 2020 December 13.9~{\sc ut}; it is not clear why the light echo of SN~1999el was not identified, though it could be related to the extended nature of the source.

\subsection{Late-time HST imaging of SN 2011ht}

\begin{figure}
\centering
\includegraphics[width=\linewidth]{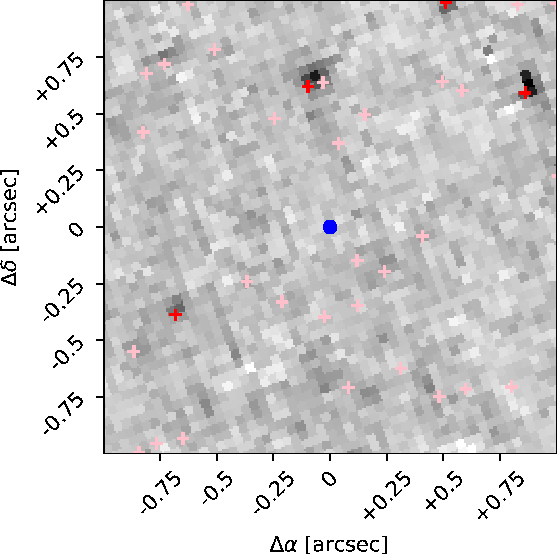}
\caption{$2 \times 2$~arcsec$^{2}$ cutout of the drizzled F555W image with north up and east left from 2017 March 10.6~{\sc ut}, centred on the position of SN2011ht (indicated with a blue circle). Sources detected at $\mathrm{S/N} > 3$ in this image are plotted in pink, while those detected with $\mathrm{S/N} > 5$ are plotted in red. A deep limit of $M_{\mathrm{F555W}} > -4.0$~mag is derived for the location of the event.}
\label{fig:hst_11ht}
\end{figure}

Based on ground-based pre-explosion non-detections of a SN~2011ht precursor with the deepest limit of $m_{g} > 22.8$~mag, \citet{roming11} excluded LBVs in hot state, and very massive $\gtrsim$30~$M_{\odot}$ YSGs and RSGs as the progenitor. We analysed two epochs of late-time HST imaging observed with WFC3, which cover the explosion site of SN~2011ht in 2017 March 10.6~{\sc ut} and 2019 April 17.1~{\sc ut}. No high resolution imaging is available of SN~2011ht, and hence an intermediate step was used to confirm the location of the event in these data. We downloaded the \textit{i}-band image covering SN~2011ht taken on 2012 February 13.5~{\sc ut}, observed with the Panoramic Survey Telescope and Rapid Response System 1 \citep[Pan-STARRS1;][]{chambers16} telescope (PS1) using the Gigapixel Camera 1, available via the PS1 Data Release 2 (DR2) archive. This image had the best seeing available in the archive among the images where SN~2011ht was still bright. 38 sources from the \textit{Gaia} DR3 that were located within 10\arcmin\ of SN~2011ht were identified, and their pixel coordinates were measured on the PS1 image. Using these pixel coordinates and their corresponding \textit{Gaia} catalogue positions, we derived a new WCS for the PS1 image. The rms uncertainty in the WCS was 0$\farcs$05 in both $\alpha$ and $\delta$. Finally, we measured the coordinates of SN~2011ht in the PS1 image, and found it to be $\alpha = 10^{\mathrm{h}}08^{\mathrm{m}}10\fs59$ and $\delta = +51\degr50\arcmin57\farcs1$, which agrees well with those reported in the TNS.

A broadly similar process was performed for the HST images. We used the pipeline-drizzled F555W image from 2019 as our reference. Only six \textit{Gaia} sources fall within the FOV, and in order to account for the higher resolution compared to the PS1 data, we corrected the catalogue positions for proper motion effects (the reference epoch of the \textit{Gaia} catalogue is 2016.0, while the WFC3 image was taken 3.3~yr later). The rms error in the new WCS found for the WFC3 image is $<$10~mas; therefore, the positional uncertainty is dominated by the position of SN~2011ht on the PS1 image. The new \textit{Gaia}-derived WCS solution was used to determine the position of SN~2011ht on the WFC3 images. No source was detected within 3$\sigma$ of this position. The {\sc dolphot} package was used to verify this, and set quantitative limits, which are reported in Table~\ref{table:hst}. The limiting magnitude is taken from the turn over of a histogram of the number of source detected with a $\mathrm{S/N} > 3$ as a function of magnitude, which indicates the loss of completeness. One source within 5$\sigma$ of the position of SN~2011ht is marginally detected in F435W at $28.0 \pm 0.3$~mag. This is quite far from the position of the SN~2011ht ($\sim$4.3~pixels) and close to the limiting magnitude for this image at $>$28.1~mag; therefore, we do not regard it as a strong associated candidate. The most constraining absolute magnitude limits for SN~2011ht are set from the 2017 images to $M_{\mathrm{F438W}} > -3.8$ and $M_{\mathrm{F555W}} > -4.0$~mag in the HST Vegamag system. A luminous and very massive surviving progenitor could be expected in the scenario of interacting consecutive non-terminal outburst shells to explain SN~1994W-like events \citep{dessart09}; however, such star is not identified at least in optical wavelengths. Unless it was obscured by newly formed dust, the limits suggest that typical supergiants should be excluded as a massive surviving precursor for SN~2011ht with non-terminal outbursts.

\subsection{Late-time HST imaging of SN 1994ak}

\begin{figure}
\centering
\includegraphics[width=\linewidth]{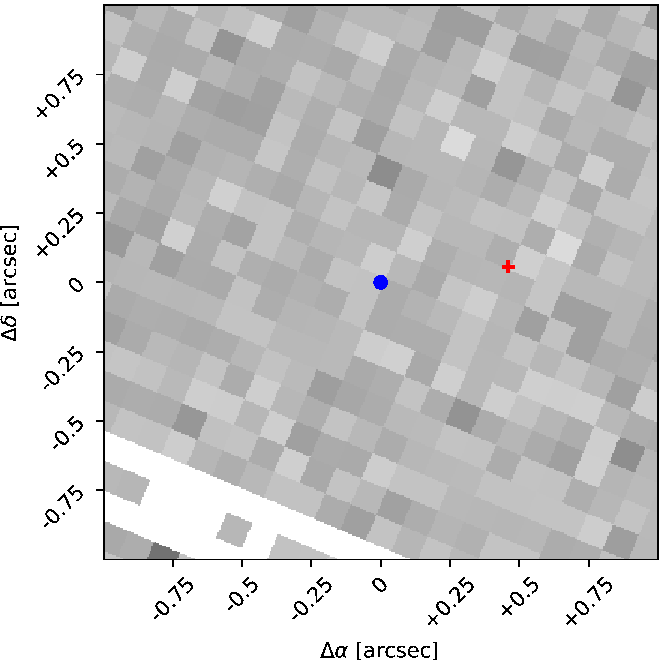}
\caption{$2 \times 2$~arcsec$^{2}$ cutout centred on the position of SN~1994ak in F555W marked with a blue circle. There are no detections by {\sc dolphot} at $\mathrm{S/N} > 3$; the marked source in red is associated with an imperfectly masked cosmic ray.}
\label{fig:hst_94ak}
\end{figure}

The explosion site of SN~1994ak was observed with WFPC2 in the F555W and F814W filters on 1997 April 18.4~{\sc ut}. The WCS in the images was verified using a single \textit{Gaia} source that fell on the detector. {\sc dolphot} was run on the images, but no source was detected within 1$\arcsec$ of the transient location, see Fig.~\ref{fig:hst_94ak}. Similar analysis was performed to the WFC3 F225W band image on 2021 September 25.1~{\sc ut}, and limiting magnitudes were derived similarly to those of SN~2011ht. Due to the larger distance and the less deep images, the photometry is not as constraining as that of SN~2011ht; however, the most stringent upper limits that were set are $M_{\mathrm{F225W}} > -7.9$ and $M_{\mathrm{F555W}} > -7.8$~mag. 

\subsection{Late-time HST imaging of SN 2004gd}

\begin{figure}
\centering
\includegraphics[width=\linewidth]{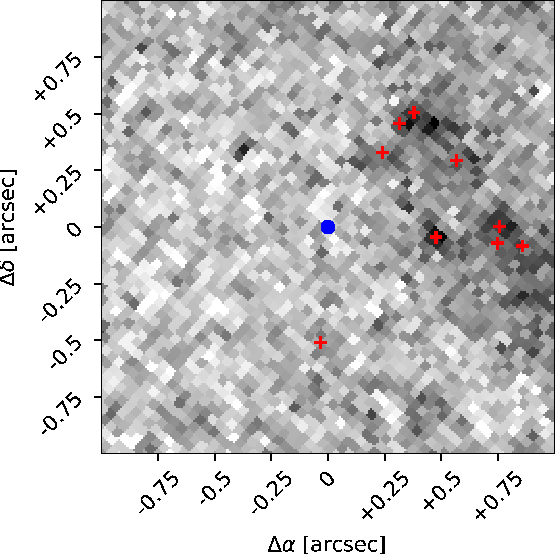}
\caption{$2 \times 2$~arcsec$^{2}$ cutout centred on the position of SN~2004gd in F225W, marked with a blue circle. Sources detected by {\sc dolphot} at $\mathrm{S/N} > 3$ are marked with red crosses.}
\label{fig:hst_04gd}
\end{figure}

WFC3 imaging was taken in the F225W filter of the site of SN~2004gd on 2013 February 10.8~{\sc ut}. Due to the paucity of UV bright sources in the field, it was not possible to align these directly to an image taken of SN~2004gd. However, we verified that the WCS in the header of the HST images was correct to one or two pixels, using four \textit{Gaia} sources in the field. Source detection was performed on the stacked, pipeline drizzled \_drc image using {\sc dolphot}, while PSF-fitting photometry of each detected source was done on the individual $\_$flc frames. {\sc dolphot} detects a number of sources in the range of $25.4 > m_{\mathrm{F225W}} > 24.8$~mag about 0$\farcs$7 to the west of where SN~2004gd exploded. As seen in Fig.~\ref{fig:hst_04gd}, it is likely that these are not genuine point sources, but rather marginally resolved extended emission from a bright star-forming region. At this distance of SN~2004gd, a 0$\farcs$7 projected separation corresponds to a physical distance of at least 250~pc. No source is seen at the position of SN~2004gd itself (nor detected by {\sc dolphot}), and we take $m_{\mathrm{F225W}} > 25.5$~mag (from the distribution of sources detected as a function of magnitude) as our $\mathrm{S/N} > 3$ upper limit, corresponding to $M_{\mathrm{F225W}} > -9.5$~mag.

\section{Discussion}
\label{sect:discussion}

By expanding the sample of SN~1994W-like transients with observations of SNe~1999eb and 2003G, it was further emphasised that these events show as common characteristic narrow spectral lines with P~Cygni minimum velocities of roughly 600 to 700~km~s$^{-1}$. We also identify SN~1999el as a SN~1994W-like event based on their spectroscopic similarity. Most of the sample events show a Type IIP SN-like light curve plateau followed by a sharp and deep drop, whereas SN~1999el has a more rapid Type IIL SN-like evolution, which could be related to the geometry of the CSM. The narrow spectral profiles of SN~1994W-like events are accompanied by broad emission wings, likely arising from electron scattering \citep{dessart09}. In the tail phase, the P~Cygni absorption components and the line wings disappear. However, no broad lines are seen in the spectroscopic observations of SNe~1994W, 2009kn, and 2011ht, which extend out to the tail phase close to 200~d after maximum light \citep{chugai04,kankare12,mauerhan13}. \citet{dessart16} carried out radiation hydrodynamical modelling of a selection of different configurations of interacting transients and concluded that the interaction of a more energetic but less massive inner ejecta with a more massive CSM shell could explain the observed evolution of SN~1994W. In the $M_{\mathrm{CSM}} > M_{\mathrm{ejecta}}$ scenario, the inner ejecta is decelerated by the reverse shock of the interaction, converting kinetic energy efficiently into radiation, reprocessed by the slowly expanding CSM shell. At the end of the plateau phase, when the system effectively becomes optically thin, followed by a sharp light curve drop, the ejecta has already been decelerated to the bulk velocity of the CSM and no broad emission lines are seen in the tail phase spectra \citep[see][]{dessart16,dessart24}. 

\begin{figure}
\centering
\includegraphics[width=\linewidth]{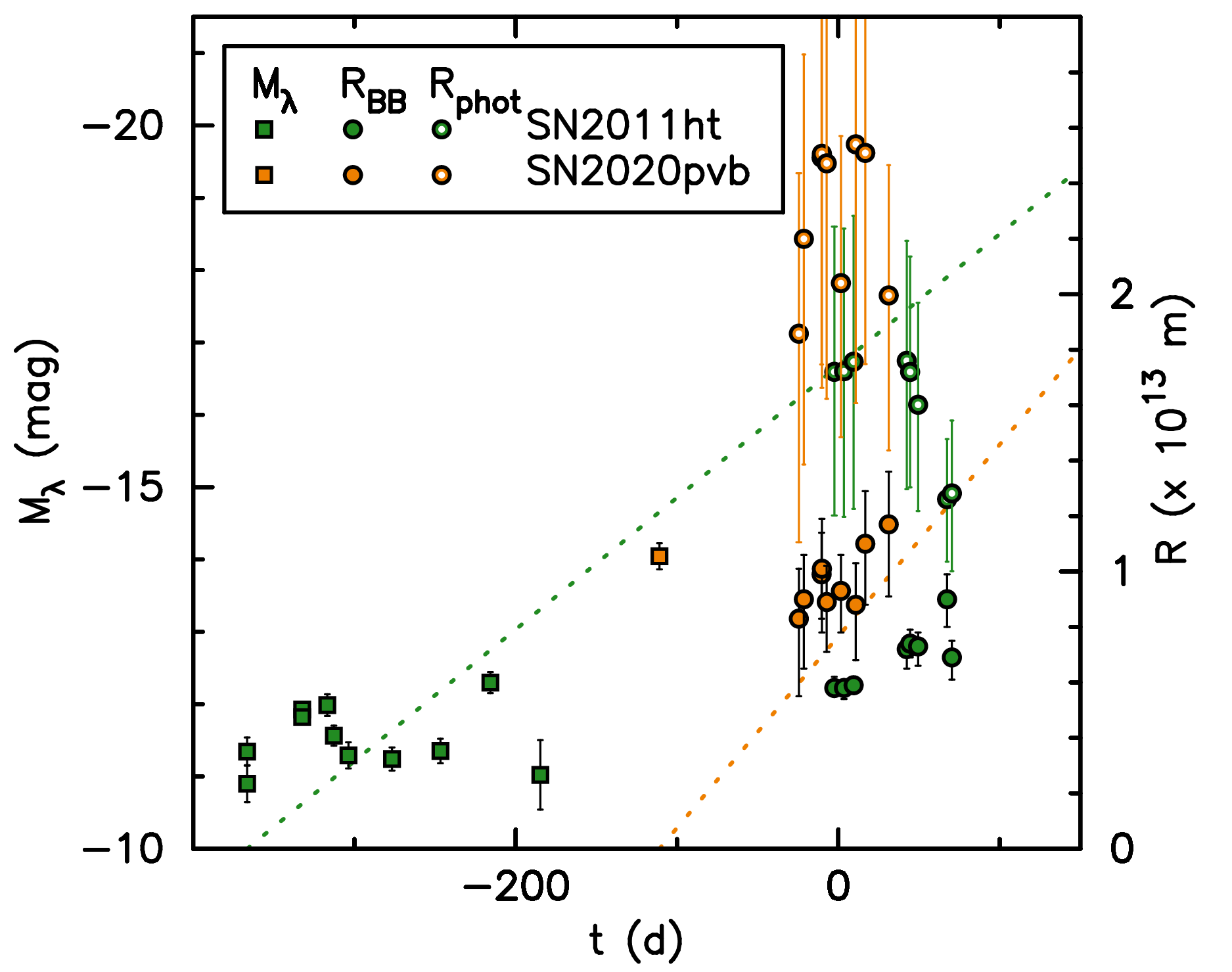}
\caption{The evolution of the ejecta radius assuming a 550~km~s$^{-1}$ bulk expansion velocity for a pre-outburst of SN~2011ht  (green dotted line) from the onset at the first epoch of detections \citep[green square symbols;][]{fraser13,ofek14} compared to the estimated $R_{\mathrm{BB}}$ and $R_{\mathrm{phot}}$ evolution of the event (green round closed and open symbols, respectively) based on spectra of the event \citep{humphreys12}. Similarly for SN~2020pvb (in pink) based on the data by \citet{elias-rosa24} assuming a velocity of 800~km~s$^{-1}$.}
\label{fig:pre-outburst}
\end{figure}

The previously reported historical outburst detections of SN~1994W-like events include the observed precursor variability before the main event of SN~2011ht \citep{fraser13,ofek14}, and one $M_{w} \approx -13.8$~mag PS1 detection of SN~2020pvb roughly 111~d before the event maximum \citep{elias-rosa24}. In Type II SNe, observed photons arise from different layers due to electron scattering in the ejecta. The photospheric radius, $R_{\mathrm{phot}}$, defined to have a total inwards integrated optical depth of $\tau = 2/3$, and the blackbody radius, $R_{\mathrm{BB}}$, fitted to the optical observations typically deviate ($R_{\mathrm{BB}} < R_{\mathrm{phot}}$); \citet{eastman96} and \citet{dessart05} have derived temperature-dependent correction factors for this effect for normal Type IIP SNe in the context of the expanding photosphere method for distance determination. However, the corrections depend on the SN characteristics, and SN~1994W-like events are not normal Type IIP SNe. \citet{dessart09} carried out detailed modelling of the observed spectra of SN~1994W and reported the $R_{\mathrm{phot}}$ values of their synthetic spectra. We derived the $R_{\mathrm{BB}}$ values for the observed spectra of SN~1994W, assuming the same parameters (e.g. distance, extinction) as adopted in \citet{dessart09}, with a blackbody fit to the spectral continuum, excluding the blue end of the spectrum ($<$4670~\AA) and wavelength regions with prominent line features. A polynomial fit to the resulting ratios of the radii yielded $R_{\mathrm{BB}}/R_{\mathrm{phot}} = 0.94 -1.45 (10^{4}/T_{\mathrm{BB}}) + 0.87 (10^{4}/T_{\mathrm{BB}})^2$. The correction error was assumed to be of the order of 0.1. We carried out a similar blackbody fit to the spectra of SNe~2011ht and 2020pvb to yield the $R_{\mathrm{BB}}$ values and used this generic correction to roughly estimate the $R_{\mathrm{phot}}$ values. These radii were compared to those expected for the material ejected in the pre-outbursts of these events. When compared to the first epoch of pre-outburst detection of SN~2011ht at $-367$~d, and assuming the characteristic velocity of 550~km~s$^{-1}$ of the event (Fig.~\ref{fig:Halpha-comp}), the kinematic age of the outburst is consistent with reaching radii within which the $R_{\mathrm{phot}}$ can be located (Fig.~\ref{fig:pre-outburst}). The expansion of the SN~2020pvb pre-outburst assuming 800~km~s$^{-1}$ is consistent within errors with the $R_{\mathrm{BB}}$, though underestimated compared to the $R_{\mathrm{phot}}$ values; if connected, either the photosphere size is overestimated, or the pre-outburst onset happened a few hundred days earlier. The deep limits before the pre-outburst detection reported by \citet{elias-rosa24} would exclude the latter scenario, unless the pre-outburst had a prominently multi-peaked light curve.

\citet{dessart16} discussed the possibility that a massive CSM shell of SN~1994W could have been produced by a nuclear flash that ejects the H envelope of a star in a RSG phase. Based on their modelling, \citet{woosley15} suggested that in the 9.0 to 10.3~$M_{\sun}$ range the nuclear flash from the silicon burning ignition could eject the loosely bound H envelope with velocities ranging from a few hundred to a few thousand km~s$^{-1}$, with luminosities ranging from roughly 10$^{39}$ to 10$^{41}$~erg~s$^{-1}$, with delay times a couple of weeks to a couple of years to the CCSN explosion, and a likely $^{56}$Ni production of 0.01 to 0.04~$M_{\sun}$. The parameter ranges are consistent with those of the SN~2011ht pre-outburst, and the generally low $^{56}$Ni mass estimates of SN~1994W-like events. Furthermore, \citet{dessart10} suggested based on numerical simulations that nuclear flashes could result in faint transients with a plateau-like light curve; the overall light curve evolution of the SN~2011ht pre-outburst could also be interpreted as an initial peak followed by a plateau with a red colour. The observed radiated energy of the pre-outburst of SN~2011ht was $\sim$10$^{46}$~erg based on the optical detections; however, to eject the H envelope the total energy of the outburst could have been notably larger due to the low efficiency of converting energy into radiation, and potentially considerable emission released in the longer IR wavelengths not probed by the optical observations. However, if H envelope ejecting nuclear flashes could be produced by stars in a certain mass range, it is perhaps unclear why the SN~1994W-like events would not be more common and are quite rare instead.

Alternatively, if SN~1994W-like transients are preceded by a contact-binary-powered LRN event, this could naturally create a massive shell of CSM via ejection of the common envelope. As an example, \citet{ercolino24} carried out computations of binary evolution models with a 12.6~$M_{\sun}$ primary, which suggested that systems with a mass ratio range of $M_{2}/M_{1} \approx 0.25$ to $0.5$ and an initial period of around 2000~d could lead to a common envelope ejection event; they also note that if such an event would happen shortly before the SN explosion, the result could be a SN~1994W-like event. Nevertheless, to have an LRN to be followed within the time scale of roughly a year by a SN explosion would likely require major fine-tuning of the system parameters. Therefore, with an emerging sample of several SN~1994W-like transients, generalising this scenario to all these events might be problematic, though some of the issues might be alleviated if the merging system consists of a massive star and a compact object (a neutron star or a black hole) where the latter could potentially spiral to the core relatively fast and result in a coalescence powered explosion \citep[see e.g.][]{barkov11,schroder20,brennan25}. 

Regardless of the uncertainties related to the proposed models of major mass-loss episodes, our NCR results for the sample of SN~1994W-like events are most consistent with relatively low-luminosity RSG progenitors and nuclear flash and stellar merger-powered events for the pre-outburst are consistent with a low-mass RSG origin. Based on the TNS database, the host galaxies of our sample of SN~1994W-like events have hosted also 4 Type Ia SNe, 8 unclassified transients, and 10 other CCSNe (6 Type II, 1 Type IIb, 2 Type Ib, and 1 Type Ibn SNe). Within the confirmed CCSNe, the relative distribution of subtypes is fairly similar to that of local volume-limited relative rates of CCSNe \citep{shivvers17}. There are some possible trends that galaxies with multiple CCSNe favour similar subtypes due to specific star formation ages \citep{anderson13}, though this appears to be a more prominent effect in highly starforming galaxies \citep{kankare21}. Furthermore, our sample suffers from small number statistics; nonetheless, these galaxies have not hosted Type Ic or other Type IIn SNe. If the hosts of SN~1994W-like transients will continue to disfavour Type Ic SNe, which are generally expected to have very massive progenitors, this would also corroborate relatively low mass progenitors for SN~1994W-like events.

\section{Conclusions}
\label{sect:conclusions}

We have investigated the properties of a compilation of spectroscopically SN~1994W-like events. Our NCR sample result of $\langle$NCR$_{\mathrm{H}\alpha}\rangle = 0.170 \pm 0.076$ for these transients is lower than a flat distribution that would follow the H$\alpha$ emission, which is a tracer of very recent star formation. This suggests that the progenitors of these events have longer lifetimes than those that trace the H$\alpha$ emission, and are therefore not very massive stars of several tens of solar masses. Instead, the NCR distribution of the sample is consistent with that of RSGs, and in particular with low-luminosity RSGs. Contrary to H$\alpha$, our sample of SN~1994W-like transients with $\langle$NCR$_{\mathrm{NUV}}\rangle = 0.488 \pm 0.084$ are consistent with a flat distribution in comparison to the host galaxy NUV emission, which is thought to trace star formation within 16 to 100~Myr, which would suggest $M_\mathrm{ZAMS} \lesssim$14~$M_{\sun}$ progenitors for the bulk of the SN~1994W-like events if rotating single stars in solar metallicity are assumed. However, with binary systems the possibilities are more complex, and somewhat larger masses cannot be excluded based on the NCR$_{\mathrm{NUV}}$ result.

We expanded the data sets of SN~1994W-like events with the analysis of SNe~1999eb and 2003G. In particular, the $-26$~d spectrum of SN~2003G shares similarities to spectra of F8-type supergiant stars and also to an example LRN NGC~4490-2011OT1; furthermore, these observations show that the early spectroscopic characteristics reported for SN~2011ht are not unique among SN~1994W-like events. The early phase spectrum of SN~2003G is also remarkably similar to the spectra of SN~1994W-like events close to the end of the plateau phase, and the spectroscopic similarity between these phases of evolution appears to be a common characteristic within SN~1994W-like events.

The late-time HST data reveals the explosion sites of SNe~1994W and 1999el to be consistent with a blended $M_{\mathrm{F814W}} = -8.6$~mag cluster and a ring-like $\sim$0$\farcs$4 diameter light echo, respectively. However, the late-time HST imaging of SN~2011ht reveals no coincident source and yields the most stringent limits for a surviving progenitor star of a SN~1994W-like event at $M_{\mathrm{F438W}} > -3.8$ and $M_{\mathrm{F555W}} > -4.0$~mag, which suggests that supergiant stars should be excluded as non-terminal precursors if no major new dust formation took place that would have obscured a massive surviving star. Less constraining limits for a surviving progenitor are set for SN~1994ak and SN~2004gd at $M_{\mathrm{F555W}} > -7.8$ and $M_{\mathrm{F225W}} > -9.5$~mag, respectively.

%%%%%%%%%%%%%%%%%%%%%%%%%%%%%%%%%%%%%%%%%%%%%%%%%%%%%%%%%%%%%%
\begin{acknowledgements}

We thank the anonymous referee for useful comments.

We thank Enrico Cappellaro, Luca Rizzi, Cristina Romero-Ca\~nizales, and Marilena Salvo for the help with some of the observations.

EK acknowledges financial support from the Emil Aaltonen foundation.

TK acknowledges support from the Research Council of Finland project 360274.

SM acknowledges support from the Research Council of Finland project 350458.

AP, NER, and SB acknowledge financial support from PRIN-INAF 2022 ``Shedding light on the nature of gap transients: from the observations to the models''.

NER acknowledges support from the Spanish Ministerio de Ciencia e Innovaci\'on (MCIN) and the Agencia Estatal de Investigaci\'on (AEI) 10.13039/501100011033 under the program Unidad de Excelencia Mar\'ia de Maeztu CEX2020-001058-M.

RK acknowledges support from the Research Council of Finland project 340613.

Based on observations made with the Nordic Optical Telescope, owned in collaboration by the University of Turku and Aarhus University, and operated jointly by Aarhus University, the University of Turku and the University of Oslo, representing Denmark, Finland and Norway, the University of Iceland and Stockholm University at the Observatorio del Roque de los Muchachos, La Palma, Spain, of the Instituto de Astrofisica de Canarias. The data presented here were obtained in part with ALFOSC, which is provided by the Instituto de Astrofisica de Andalucia (IAA) under a joint agreement with the University of Copenhagen and NOT.

GALEX is operated for NASA by the California Institute of Technology under NASA contract NAS5-98034. 

Based on observations made with ESO Telescopes at the La Silla Paranal Observatory under programme IDs 67.D-0594 and 164.H-0376.

Based on observations collected at Copernico telescope (Asiago, Italy) of the INAF - Osservatorio Astronomico di Padova.

Based on observations made with the WHT and INT operated on the island of La Palma by the Isaac Newton Group of Telescopes in the Spanish Observatorio del Roque de los Muchachos of the Instituto de Astrof\'isica de Canarias. This paper makes use of data obtained from the Isaac Newton Group of Telescopes Archive which is maintained as part of the CASU Astronomical Data Centre at the Institute of Astronomy, Cambridge.

This research is based on observations made with the NASA/ESA Hubble Space Telescope obtained from the Space Telescope Science Institute, which is operated by the Association of Universities for Research in Astronomy, Inc., under NASA contract NAS 5-26555. These observations are associated with programs 6673, 8602, 9042, 9788, 12229, 13007, 14149, 14614, 14668, 15166, 15645, 16179, 16241, 16691, and 17070.

The Pan-STARRS1 Surveys (PS1) and the PS1 public science archive have been made possible through contributions by the Institute for Astronomy, the University of Hawaii, the Pan-STARRS Project Office, the Max-Planck Society and its participating institutes, the Max Planck Institute for Astronomy, Heidelberg and the Max Planck Institute for Extraterrestrial Physics, Garching, The Johns Hopkins University, Durham University, the University of Edinburgh, the Queen's University Belfast, the Harvard-Smithsonian Center for Astrophysics, the Las Cumbres Observatory Global Telescope Network Incorporated, the National Central University of Taiwan, the Space Telescope Science Institute, the National Aeronautics and Space Administration under Grant No. NNX08AR22G issued through the Planetary Science Division of the NASA Science Mission Directorate, the National Science Foundation Grant No. AST-1238877, the University of Maryland, Eotvos Lorand University (ELTE), the Los Alamos National Laboratory, and the Gordon and Betty Moore Foundation.

This research has made use of the NASA/IPAC Extragalactic Database (NED) which is operated by the Jet Propulsion Laboratory, California Institute of Technology, under contract with the National Aeronautics and Space Administration. 

This publication makes use of data products from the Two Micron All Sky Survey, which is a joint project of the University of Massachusetts and the Infrared Processing and Analysis Center/California Institute of Technology, funded by the National Aeronautics and Space Administration and the National Science Foundation.

Funding for SDSS-III has been provided by the Alfred P. Sloan Foundation, the Participating Institutions, the National Science Foundation, and the U.S. Department of Energy Office of Science. The SDSS-III web site is \urlwofont{http://www.sdss3.org/}. SDSS-III is managed by the Astrophysical Research Consortium for the Participating Institutions of the SDSS-III Collaboration including the University of Arizona, the Brazilian Participation Group, Brookhaven National Laboratory, Carnegie Mellon University, University of Florida, the French Participation Group, the German Participation Group, Harvard University, the Instituto de Astrofisica de Canarias, the Michigan State/Notre Dame/JINA Participation Group, Johns Hopkins University, Lawrence Berkeley National Laboratory, Max Planck Institute for Astrophysics, Max Planck Institute for Extraterrestrial Physics, New Mexico State University, New York University, Ohio State University, Pennsylvania State University, University of Portsmouth, Princeton University, the Spanish Participation Group, University of Tokyo, University of Utah, Vanderbilt University, University of Virginia, University of Washington, and Yale University. 

{\sc gaia} is a derivative of the Skycat catalogue and image display tool, developed as part of the VLT project at ESO. Skycat and {\sc gaia} are free software under the terms of the GNU copyright. 

We have made use of the Weizmann interactive supernova data repository (\urlwofont{www.weizmann.ac.il/astrophysics/wiserep}).

We have made use of the Ned Wright's cosmology calculator \citep[][\urlwofont{www.astro.ucla.edu/~wright/CosmoCalc.html}]{wright06}.
\end{acknowledgements}

\begin{appendix}

\onecolumn
\section{Additional tables and figures}

\begin{table*}[h!]
\caption{New photometry for SN~1999eb with the errors given in brackets.}
\centering
\small
\setlength\tabcolsep{3.5pt}
\begin{tabular}{cccccccccl}
\hline
\hline
JD & $m_{U}$ & $m_{B}$ & $m_{V}$ & $m_{R}$ & $m_{I}$ & $m_{J}$ & $m_{H}$ & $m_{K}$ & Telescope \\
 & (mag) & (mag) & (mag) & (mag) & (mag) & (mag) & (mag) & (mag) & \\ 
\hline
2451456.48 & - & - & - & - & - & 15.78(0.06) & 15.67(0.06) & 15.44(0.10) & TNG/ARNICA \\ 
2451457.51 & - & - & - & - & - & 15.67(0.08) & 15.50(0.06) & 15.26(0.09) & TNG/ARNICA \\ 
2451457.59 & 15.74(0.02) & 16.42(0.02) & 16.22(0.03) & 16.02(0.05) & 15.94(0.04) & - & - & - & TNG/OIG \\ 
2451458.39 & - & 16.38(0.02) & 16.22(0.02) & 15.99(0.03) & - & - & - & - & AS1.82/AFOSC \\ 
2451458.56 & 15.72(0.04) & 16.39(0.02) & 16.22(0.04) & 16.01(0.05) & 15.91(0.04) & - & - & - & TNG/OIG \\ 
2451459.41 & - & 16.31(0.03) & 16.20(0.01) & 16.00(0.03) & 15.91(0.04) & - & - & - & AS1.82/AFOSC \\ 
2451459.50 & 15.69(0.05) & 16.36(0.01) & 16.22(0.02) & 15.96(0.05) & 15.89(0.03) & - & - & - & TNG/OIG \\ 
2451459.66 & 15.67(0.05) & 16.31(0.03) & 16.17(0.01) & 15.97(0.03) & 15.87(0.04) & - & - & - & ESO3.6/EFOSC2 \\ 
2451460.67 & - & - & - & - & - & 15.66(0.04) & 15.42(0.04) & 15.24(0.04) & VLT/ISAAC \\ 
2451461.51 & - & 16.22(0.03) & 16.11(0.03) & 15.98(0.05) & 15.88(0.05) & - & - & - & AS1.82/AFOSC \\ 
2451465.51 & 15.56(0.04) & - & 16.00(0.03) & 15.84(0.05) & 15.83(0.04) & - & - & - & TNG/OIG \\ 
2451472.61 & 15.51(0.05) & 16.20(0.02) & 15.99(0.01) & 15.81(0.02) & 15.70(0.02) & - & - & - & D1.54/DFOSC \\ 
2451485.45 & 15.61(0.05) & 16.29(0.01) & 16.01(0.03) & 15.83(0.05) & 15.79(0.04) & - & - & - & TNG/OIG \\ 
2451485.70 & - & 16.28(0.02) & 16.03(0.01) & 15.83(0.02) & - & - & - & - & D1.54/DFOSC \\ 
2451486.66 & - & 16.29(0.02) & 16.02(0.01) & 15.85(0.02) & 15.78(0.03) & - & - & - & D1.54/DFOSC \\ 
2451486.68 & - & - & - & - & - & 15.55(0.12) & 15.52(0.32) & 15.38(0.61) & NTT/SOFI \\ 
2451517.66 & 16.14(0.05) & 16.67(0.02) & 16.39(0.03) & 16.17(0.05) & 15.90(0.04) & - & - & - & ESO3.6/EFOSC2 \\ 
2451520.40 & - & 16.68(0.03) & 16.35(0.03) & 16.13(0.02) & 15.83(0.15) & - & - & - & AS1.82/AFOSC \\ 
2451525.40 & 16.37(0.05) & 16.77(0.02) & 16.39(0.03) & 16.21(0.05) & 16.03(0.04) & - & - & - & TNG/OIG \\ 
2451530.61 & 16.51(0.08) & 16.86(0.05) & 16.56(0.15) & 16.28(0.16) & 15.82(0.20) & - & - & - & ESO3.6/EFOSC2 \\ 
2451541.57 & - & - & 16.57(0.01) & 16.33(0.02) & 16.10(0.03) & - & - & - & D1.54/DFOSC \\ 
2451782.69 & - & 20.86(0.14) & 19.96(0.07) & 19.34(0.16) & - & - & - & - & TNG/OIG \\ 
\hline
\end{tabular}
\label{table:phot_99eb}
\end{table*}

\begin{table*}[h!]
\caption{Spectroscopic log of observations for SN 1999eb.}
\centering
\small
\setlength\tabcolsep{2.5pt}
\begin{tabular}{ccccccl}
\hline
\hline
JD & $t$ & Grism/grating & Slit & $R$ & $t_{\mathrm{integration}}$ & Telescope \\
 & (d) & & ($''$) & & (s) & \\ 
\hline
2451458.5 & $-17$ & GR04 & 2.1 & 290 & 1800 & AS1.82/AFOSC \\
2451459.5 & $-16$ & GR04 & 2.1 & 290 & 2800+2700 & AS1.82/AFOSC \\
2451459.7 & $-16$ & Gr\#11 & 1.5 & 230 & 600 & ESO3.6/EFOSC2 \\
2451469.7 & $-6$ & Grat\#5 & 1.0 & 400 & 2x600 & NTT/EMMI \\
2451482.5 & $+7$ & R150V & 1.5 & 330 & 4x1800 & INT/IDS \\
2451485.7 & $+10$ & Gr\#4 & 1.5 & 470 & 3600 & D1.54/DFOSC \\
2451486.7 & $+11$ & Gr\#4 & 1.5 & 470 & 3600 & D1.54/DFOSC \\
2451489.7 & $+24$ & R300B, R316R & 1.0 & 1100, 1800 & 1200 & WHT/ISIS \\
2451517.6 & $+42$ & Gr\#11 & 1.5 & 230 & 1200 & ESO3.6/EFOSC2 \\
2451541.6 & $+66$ & Gr\#4 & 2.0 & 470 & 1800 & D1.54/DFOSC \\
\hline
\end{tabular}
\label{table:spect_99eb}
\end{table*}

\begin{table*}[h!]
\caption{Late-time HST photometry of the explosion sites of SN~1994W-like events with the errors given in brackets.}
\centering
\small
\setlength\tabcolsep{2.5pt}
\begin{tabular}{lccccccccccl}
\hline
\hline
SN & Date & JD & $m_{\mathrm{F225W}}$ & $m_{\mathrm{F336W}}$ & $m_{\mathrm{F438W}}$ & $m_{\mathrm{F450W}}$ & $m_{\mathrm{F555W}}$ & $m_{\mathrm{F606W}}$ & $m_{\mathrm{F658N}}$ & $m_{\mathrm{F814W}}$ & Instrument \\
 & (UT) & & (mag) & (mag) & (mag) & (mag) & (mag) & (mag) & (mag) & (mag) & \\
\hline
1994W & 2001-07-04.8 & 2452095.3 & - & - & - & 22.76(0.06) & - & 22.99(0.05) & - & 23.53(0.18) & WFPC2 \\
1994W & 2004-05-29.6 & 2453155.1 & - & - & - & - & - & - & 22.58(0.18) & 23.34(0.11) & ACS \\
1994W & 2011-01-30.0 & 2455591.5 & - & 22.43(0.11) & - & - & - & - & - & - & WFC3 \\
1994ak & 1997-04-18.4 & 2450556.9 & - & - & - & - & $>$25.3 & - & - & $>$24.3 & WFPC2 \\
1994ak & 2021-09-25.1 & 2459482.6 & $>$25.3 & - & - & - & - & - & - & - & WFC3 \\
1999el & 2003-10-24.5 & 2452937.0 & - & - & - & - & - & - & - & 23.64(0.17) & ACS \\
1999el & 2015-11-25.6 & 2457352.1 & - & - & - & - & 24.13(0.22) & - & - & 23.89(0.23) & WFC3 \\
1999el & 2016-10-15.0 & 2457676.5 & - & - & - & - & 24.11(0.28) & - & - & 23.70(0.18) & WFC3 \\
1999el & 2017-11-10.8 & 2458068.3 & - & - & - & - & 24.23(0.25) & - & - & 23.75(0.25) & WFC3 \\
1999el & 2019-05-23.0 & 2458626.5 & - & - & - & - & - & - & - & 24.03(0.30) & ACS \\
1999el & 2020-12-13.9 & 2459197.4 & - & - & - & - & 24.33(0.21) & - & - & 23.95(0.27) & WFC3 \\
1999el & 2022-02-17.1 & 2459627.6 & - & - & - & - & 24.40(0.25) & - & - & 23.96(0.23) & WFC3 \\
1999el & 2023-02-22.3 & 2459997.8 & - & - & - & - & 24.33(0.30) & - & - & 24.07(0.29) & ACS \\
2004gd & 2013-02-10.8 & 2456334.3 & $>$25.5 & - & - & - & - & - & - & - & WFC3 \\
2011ht & 2017-03-10.6 & 2457823.1 & $>$26.5 & - & $>$28.1 & - & $>$27.9 & - & - & $>$27.0 & WFC3 \\
2011ht & 2019-04-17.1 & 2458590.6 & - & - & - & - & $>$27.4 & - & - & $>$26.3 & WFC3 \\
\hline
\end{tabular}
\label{table:hst}
\end{table*}

\begin{figure*}[h!]
\centering
\includegraphics[width=0.49\linewidth]{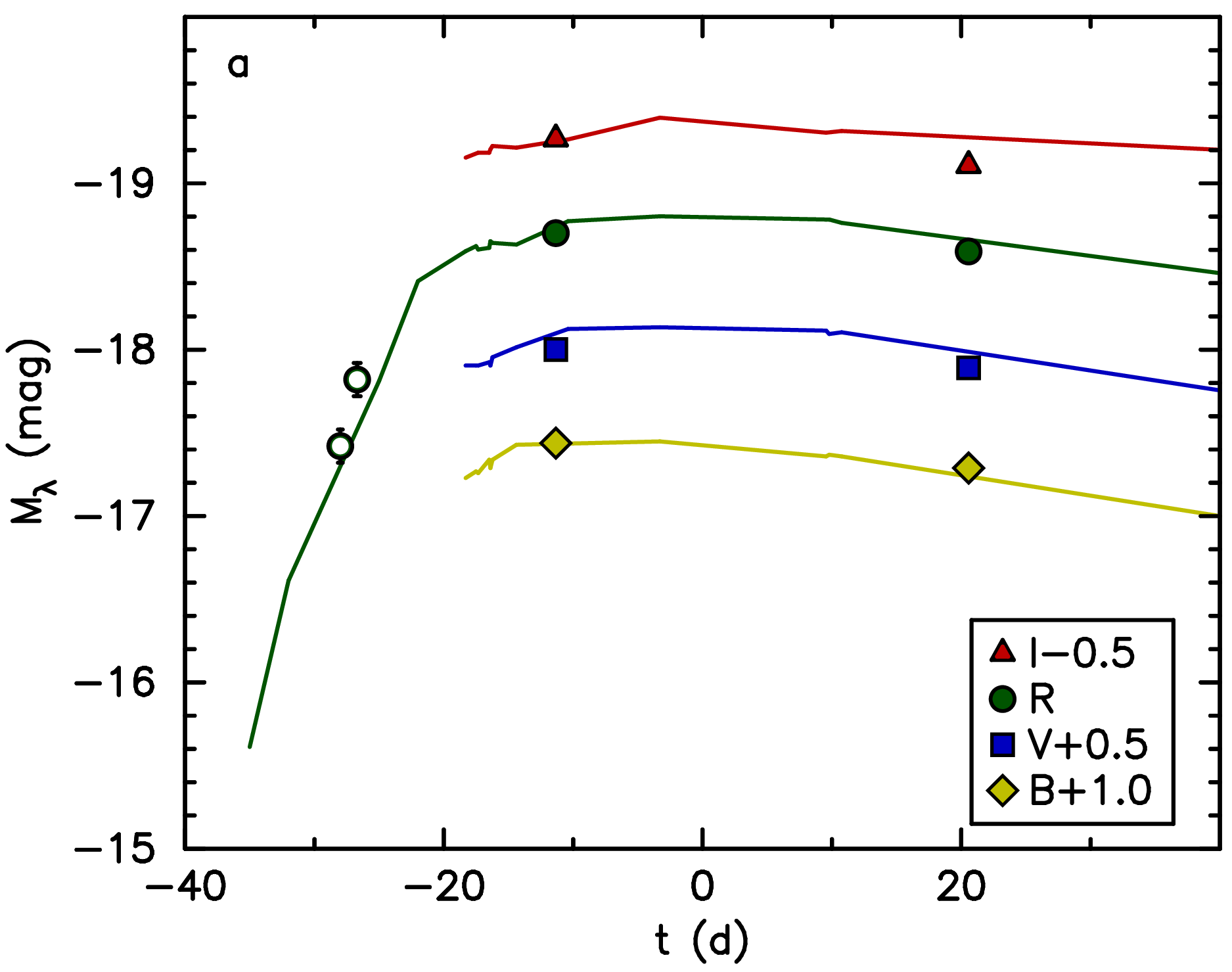}
\includegraphics[width=0.49\linewidth]{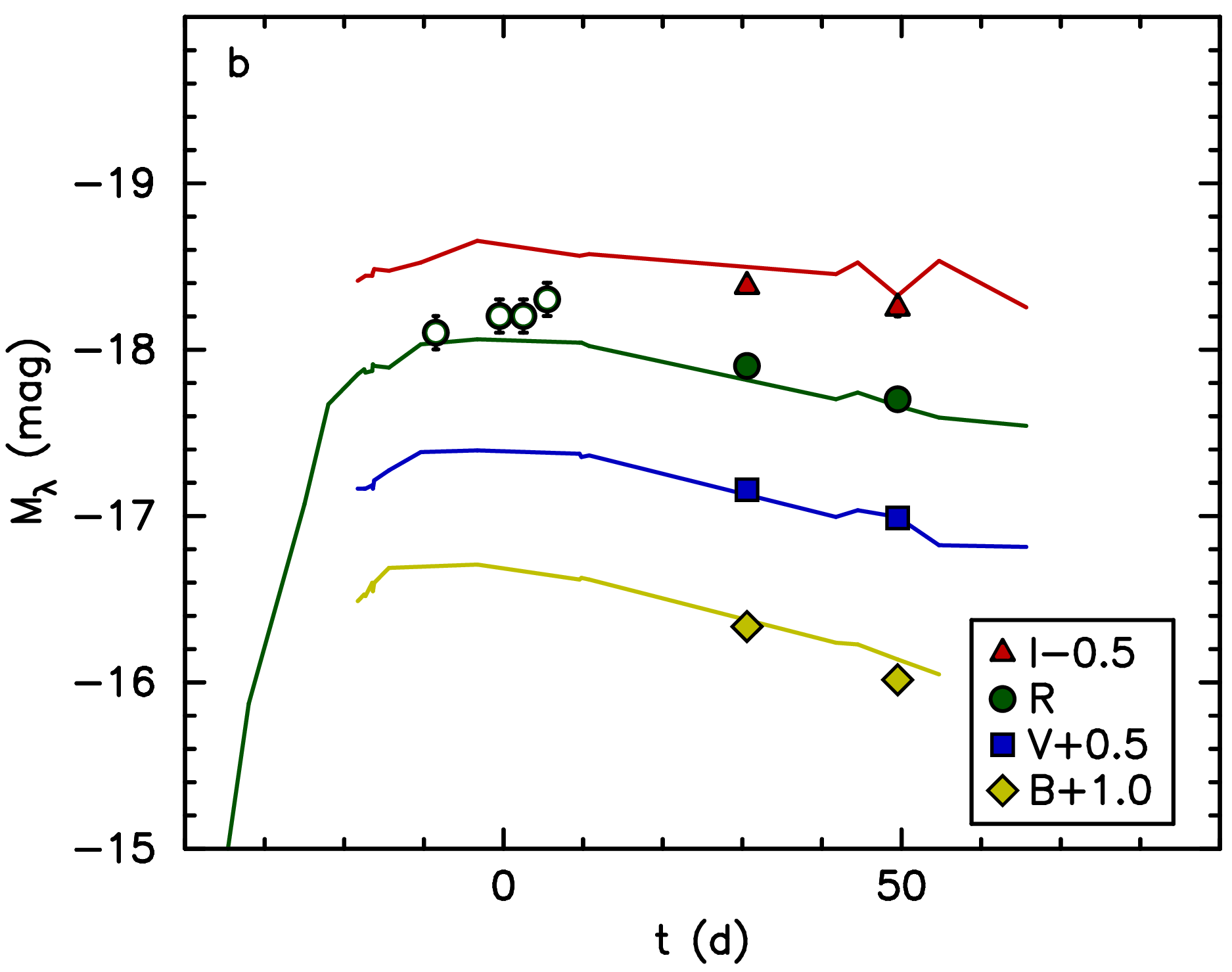}
\caption{a) Absolute light curves of SN~2003G (points) with the estimated negligible host galaxy extinction adopted. The early observations by \citet{graham03} are shown with open symbols. The comparison light curves (solid lines) of SN~1999eb (Sect.~\ref{sect:99eb}) are shown vertically shifted by $-0.1$~mag. b) Absolute light curves of SN~2004gd (points) with the estimated host galaxy extinction of $A_{V} = 1.0$~mag; the photometry of the event was carried out with Pan-STARRS1 DR2 images as rough templates. The early observations by \citet{moore04} are shown with open symbols. The comparison light curves (solid lines) of SN~1999eb are shown vertically shifted by $+0.6$~mag.}
\label{fig:03G}
\end{figure*}

\begin{figure*}[h!]
\centering
\includegraphics[width=0.49\linewidth]{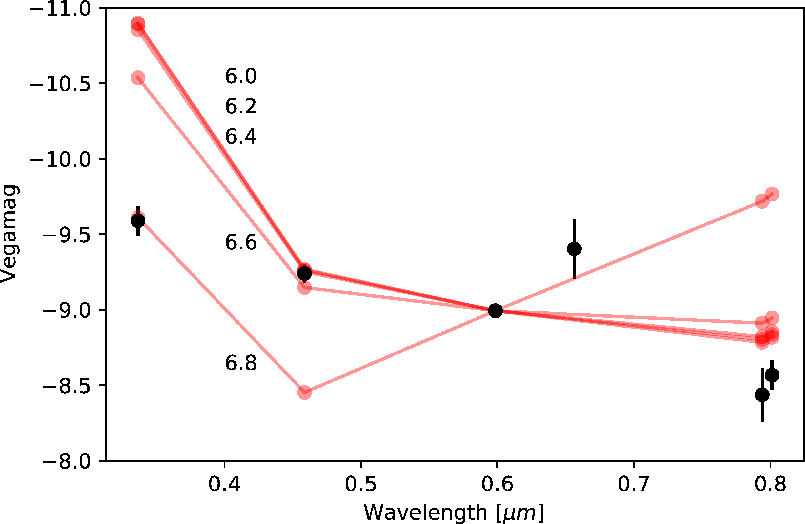}
\caption{SED fits to photometry of the putative host cluster of SN~1994W. The combined magnitude of both the point source and any diffuse emission is shown at the wavelength of each band. SED models have been scaled to match the F606W band, and are plotted in red. We see that populations with log(Age) between 6.0 and 6.6~dex match, with the exception of the F336W band, while the older 6.8 dex population does not match either the F450W or the F814W band.}
\label{fig:cluster_94W}
\end{figure*}

\end{appendix}
\end{document}